\documentclass[aps,prd,tightenlines,notitlepage,preprintnumbers,superscriptaddress,nofootinbib,11pt]{revtex4-2}
\usepackage[utf8]{inputenc}
\usepackage[letterpaper, total={7in, 9.3in}, top=1in]{geometry}
\usepackage{amsmath}
\usepackage{amssymb}
\usepackage{enumerate}
\usepackage{amsmath}
\usepackage{tikz}
\usepackage{cancel}
\usepackage[compat=1.1.0]{tikz-feynman}
\usepackage[caption=false]{subfig}
\usepackage{feynmf}
\usepackage{slashed}
\usepackage{braket}
\usepackage{url}
\usepackage{natbib}
\usepackage{graphicx}
\usepackage[pdftex,bookmarks,linktocpage,pdfpagelabels,plainpages=false,hyperfigures,linkcolor=teal,citecolor=teal,urlcolor=teal]{hyperref} %for hypertext links
\hypersetup{colorlinks=true}

\usepackage{mathrsfs,amssymb}  
\usepackage{cancel}
\usepackage[normalem]{ulem}

\usepackage{makecell}
\usepackage{tabularx}
\newcolumntype{R}{>{\raggedleft\arraybackslash}X}
\usepackage{ragged2e}
\usepackage[capitalize]{cleveref}

\usepackage{orcidlink}
\usepackage{fontawesome}

\tikzfeynmanset{
  fermion1/.style={
    /tikz/postaction={
      /tikz/decoration={
        markings,
        mark=at position 0.5 with {
          \node[
            transform shape,
            xshift=-0.25mm,
            fill,
            dart tail angle=100,
            inner sep=0.65pt,
            draw=none,
            dart
          ] { };
        },
      },
      /tikz/decorate=true,
    },
  },
  fermion2/.style={
    /tikz/postaction={
      /tikz/decoration={
        markings,
        mark=at position 0.5 with {
          \arrow{>[length=6pt, width=5pt]};
        },
      },
      /tikz/decorate=true,
    },
  },
}

\makeatletter
\def\@hangfrom@section#1#2#3{\@hangfrom{#1#2}#3}%\MakeTextUppercase{#3}}%
\def\@hangfroms@section#1#2{#1#2}%\MakeTextUppercase{#2}}%
\makeatother

\usepackage{xcolor}
\usepackage{xspace}

\newcommand{\gitlink}{\href{https://github.com/athompson-git/alplib}{\textsc{g}it\textsc{h}ub~{\large\color{black}\faGithub}}\xspace}

\begin{document}

\begin{flushright}
MI-HET-862\\
CETUP2025-007
\end{flushright}

\title{Finding BSM Needles in Electromagnetic Haystacks at DUNE}

\author{Vedran Brdar \orcidlink{0000-0001-7027-5104}}
\email{vedran.brdar@okstate.edu}
\affiliation{
Department of Physics, Oklahoma State University, Stillwater, OK, 74078, USA}

\author{Bhaskar Dutta \orcidlink{0000-0002-0192-8885}}
\email{dutta@tamu.edu}
\affiliation{Texas~A$\&$M~University,~College~Station,~TX~77843,~USA}

\author{Wooyoung Jang \orcidlink{0000-0001-9413-8948}}
\email{wooyoung.jang@uta.edu}
\affiliation{Department of Physics, University of Texas, Arlington, TX 76019, USA}

\author{Doojin Kim \orcidlink{0000-0002-4186-4265}}
\email{doojin.kim@usd.edu }
\affiliation{Texas~A$\&$M~University,~College~Station,~TX~77843,~USA}
\affiliation{Department of Physics, University of South Dakota, Vermillion, SD 57069, USA}

\author{Ian~M.~Shoemaker \orcidlink{0000-0001-5434-3744}}
\email{shoemaker@vt.edu}
\affiliation{Center for Neutrino Physics, Department of Physics, Virginia Tech, Blacksburg, VA 24061, USA}

\author{Zahra Tabrizi \orcidlink{0000-0002-0592-7425}}
\email{z$\_$tabrizi@pitt.edu}
\affiliation{PITT PACC, Department of Physics and Astronomy, University of Pittsburgh, 3941 O’Hara St., Pittsburgh, PA 15260, USA}
\affiliation{Theoretical Physics Department, CERN, 1 Esplanade des Particules, CH-1211 Geneva 23, Switzerland}

\author{Adrian Thompson \orcidlink{0000-0002-9235-0846}}
\email{a.thompson@northwestern.edu}
\affiliation{Texas~A$\&$M~University,~College~Station,~TX~77843,~USA}
\affiliation{Northwestern~University,~Evanston,~IL~60208,~USA}

\author{Jaehoon~Yu \orcidlink{0000-0001-7632-0033}}
\email{jaehoonyu@uta.edu}
\affiliation{Department of Physics, University of Texas, Arlington, TX 76019, USA}

%%%%%%%%%%%%%%%%%%%%%%%%%%%%%%%%%%%%%%%%%%%%%%%%%%%%%%%%%%%%%%%%%%%%%%%%%%%%%
\begin{abstract}
In this work, motivated by several beyond the Standard Model signal topologies, we perform detailed background mitigation analyses for the DUNE near detector. Specifically, we investigate $e^+ e^-$, $e^- \gamma$, $\gamma$, and $\gamma\gamma$ final states that may arise from long-lived particles, including light mediators, dark matter, heavy neutral leptons, and axion-like particles (ALPs), decaying or scattering inside the liquid argon detector. 
To this end, we employ both photophilic and leptophilic ALPs as phenomenological benchmarks. The aforementioned final states leave a hard electromagnetic signature with no hadronic activity above the detector energy thresholds. Nevertheless, such signatures are not immune to backgrounds from neutrino scattering in the detector, which are in the focus of our study. In order to model realistic experimental analyses, we take into account particle misidentification rates, cross-contamination effects, and detector responses. We calculate confidence limit projections for DUNE, thereby presenting realistic capabilities for constraining or discovering new physics manifested through electromagnetic showers.
\end{abstract}
%%%%%%%%%%%%%%%%%%%%%%%%%%%%%%%%%%%%%%%%%%%%%%%%%%%%%%%%%%%%%%%%%%%%%%%%%%%%%%%%

\maketitle

\section{Introduction}

While the DUNE \cite{DUNE:2020txw} experiment is expected to primarily address the present unknowns in neutrino oscillation physics, it will also offer great opportunities for constraining and discovering various beyond-the-Standard-Model (BSM) realizations \cite{DUNE:2022aul}. Regarding near-detector (ND) opportunities, such BSM targets include, but are not limited to, neutrino interactions through higher dimensional operators, searches for light sterile neutrinos, heavy neutral leptons, light dark matter, new mediators, anomalous tau neutrino appearance, as well as searches for light fermions and bosons (see e.g. ref.~\cite{Batell:2022xau} and references therein).
In connection to the new mediator searches, one of the well-motivated new physics scenarios are light pseudoscalars that may originate as the pseudo-Nambu-Goldstone bosons of a spontaneously broken global symmetry. A prominent example is the axion, which was originally proposed to address the strong CP problem~\cite{Peccei:1977hh,Wilczek:1977pj,Weinberg:1977ma}, and which may also provide an explanation for dark matter~\cite{Preskill:1982cy,Abbott:1982af,Dine:1982ah,Duffy:2009ig,Marsh:2015xka,Battaglieri:2017aum} or the observed matter-antimatter asymmetry of the Universe~\cite{Co:2019wyp,Co:2020jtv,Im:2021xoy}.
Traditional QCD axion models~\cite{Dine:1981rt,Zhitnitsky:1980tq,Kim:1979if,Shifman:1979if} face stringent bounds from astrophysics and terrestrial experiments, especially for masses above the eV scale, strongly motivating searches for light axions as solutions to the dark matter problem and matter-antimatter asymmetry. Axion solutions to the smallness of strong CP violation are not without complications from ultra-violet physics; it is conjectured that gravity (or other new physics at lower scales) breaks exact global symmetries, in this case through Planck-suppressed operators that may spoil the axion potential's ability to minimize strong CP violation; the so-called \textit{quality problem}. Heavy QCD axions in less-minimal extensions have been put forth with this motivation in mind, exploring the role of small instanton physics or enlarged color group extensions of the SM~\cite{PhysRevD.93.115010,Gaillard:2018xgk,PhysRevLett.124.221801,Kivel:2022emq,PhysRevD.106.015030,Kitano:2021fdl,Valenti:2022tsc}, though enlarged axion masses raise additional theoretical questions~\cite{Dine:2022mjw}. New pseudoscalar axion-like particles (ALPs) that do not necessarily solve the strong CP problem can also be motivated within string theory frameworks (e.g., ref.~\cite{Cicoli:2012sz}), and others can help explain the smallness of the electroweak scale through \textit{relaxion} implementations that have broad mass ranges including above the keV scale~\cite{Graham:2015cka,Banerjee:2020kww}; in such a scenario, the baryon-antibaryon asymmetry through heavy \textit{axiogenesis} mechanism can be achieved~\cite{Co:2022aav}.

In many of these scenarios, dominant couplings to either photons, electrons, or both can be realized. Simple examples include variations of the traditional KSVZ and DFSZ axion, generating dominant photon or electron couplings, respectively, and broadly in other contexts~\cite{DiLuzio:2020wdo,Cicoli:2012sz}. Accelerator-based searches at neutrino experiments for long-lived ALPs decaying to electromagnetic final states can also be particularly challenging from the standpoint of background discrimination, motivating a phenomenological study. In this work, we will consider heavy,~$\gtrsim$~keV-mass ALPs, especially interacting with the SM photon and electron, as a case study; such a particle has rather similar properties to the QCD axion, with the exception that the particle's mass and decay constant are considered as independent parameters. Various experimental efforts have been proposed to search for axions and ALPs using their couplings with the SM particles, see e.g., \cite{Fortin:2021cog,Adams:2022pbo} and references therein.

Regarding the opportunities for ALP searches at DUNE, let us point out that the sensitivity of the DUNE ND to di-photon final states was pioneered in our earlier work~\cite{Brdar:2020dpr}; it was shown that an ALP flux component could arise from the Primakoff scattering process of secondary photons produced by neutral meson decays and electromagnetic showers in the DUNE target. Other ALP production mechanisms inside the beam target are possible as well, e.g., by utilizing electron couplings \cite{Capozzi:2023ffu,CCM:2021jmk}, couplings to the electroweak bosons~\cite{Coloma:2023oxx}, kaon portals~\cite{Berger:2024xqk}, and nuclear couplings from gluon dominance~\cite{Kelly:2020dda}.
In this work, which builds upon \cite{Brdar:2020dpr,Brdar:2022vum, Capozzi:2023ffu}, we will focus on background events that could conceal a signal from the decays or scattering of a long-lived ALP. Our background study is also directly relevant to other BSM physics giving rise to the same final states, including other bosonic mediators, heavy neutral leptons, and light dark matter. This possibility is especially interesting in the environments of liquid argon time projection chambers (LArTPCs), including the DUNE ND and the ongoing short-baseline neutrino program at Fermilab \cite{Machado:2019oxb}, namely SBND, MicroBooNE, and ICARUS detectors.
We will present our findings for the capability for particle identification and momentum reconstruction of charged particle tracks which would allow for powerful separation of signal from background. Note that in the recent related study, ref.~\cite{Coloma:2023oxx}, the authors considered ALP searches with $\mu^+ \mu^-$, $e^+ e^-$, $\gamma\gamma$, and $\pi^0 \pi^+ \pi^-$ final states from the decays of a hadronically-produced ALP, and found that a number of kinematic cuts could almost completely reduce the relevant backgrounds in these final state topologies. In this work, we show that ALPs produced from electromagnetic secondary particles in the beam target and leaving $e^+ e^-$, $e^- \gamma$, $\gamma$, and $\gamma\gamma$ from ALP decays (including collinear decays merged into single-showers) and scattering signatures in the detector have reducible backgrounds.

The paper is organized as follows. In \cref{sec:model}, we outline the considered ALP model and enumerate its relevant detection channels and final-state particle topologies. In \cref{sec:sims}, we present the simulation results for both the target (ALP production) and detector. \Cref{sec:selection} is dedicated to the preselection cuts and selection cuts for each of the considered signal topologies. In \cref{sec:sensitivity} we calculate sensitivity projections for DUNE ND-LAr by leveraging background mitigation strategies developed in the form of selection cuts. We also compare results with and without backgrounds in order to illustrate the importance of understanding neutrino-nucleus interactions (hence, neutrino-induced backgrounds) in this type of new physics searches. Finally, in \cref{sec:conclusion} we conclude our findings.
%%%%%%%%%%%%%%%%%%%%%%%%%%%%%%%%%%%%%%%%%%%%%%%%%%%%%%%%%%%%%%%%%%%%%%%%%%%%%%%%%%%

%%%%%%%%%%%%%%%%%%%%%%%%%%%%%%%%%%%%%%%%%%%%%%%%%%%%%%%%%%%%%%%%%%%%%%%%%
\section{Phenomenological Model and Preliminaries}\label{sec:model}
\begin{figure}[tbh]
 \centering
  %%% PHOTON SCATTERING (INV PRIMAKOFF)
    \includegraphics[width=1.0\textwidth]{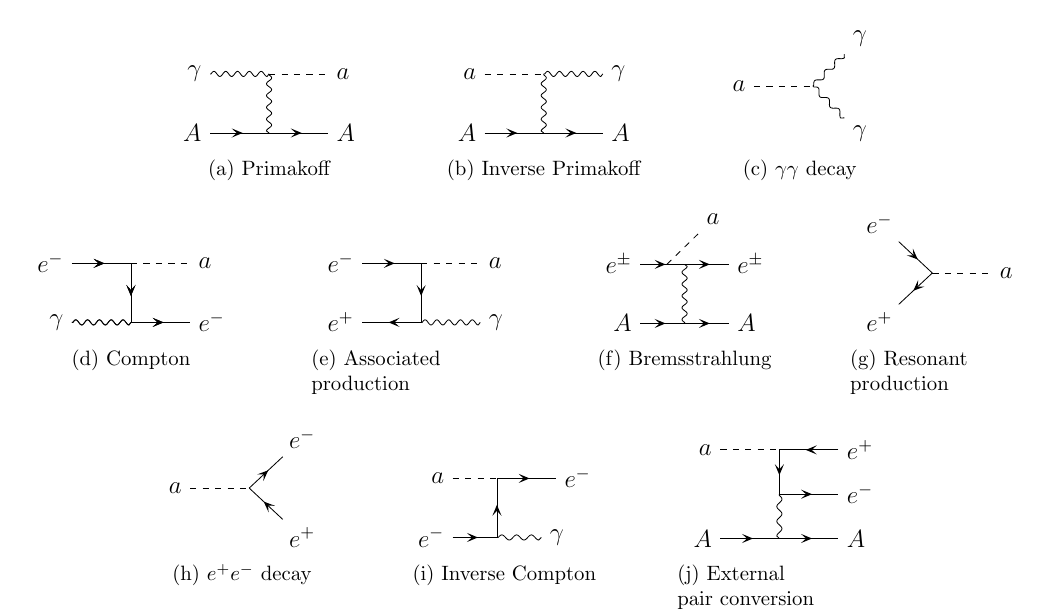}

    \caption{Top row: The $g_{a\gamma}$-driven (a) ALP production from Primakoff scattering and detection (b,c) due to inverse Primakoff and di-photon decays. Middle row: $g_{ae}$-driven ALP production from Compton scattering, associated production from electron-positron annihilation, bremsstrahlung, and resonant annihilation. Bottom row: $g_{ae}$-driven detection through decays to an electron-positron pair, inverse Compton scattering, and electron-positron pair production through atomic scattering.}
    \label{fig:feynman}
\end{figure}

To benchmark the expectations for DUNE sensitivity for BSM searches with the inclusion of backgrounds, we consider a phenomenological model featuring an ALP coupling to electrons and photons:
\begin{equation}
\label{eq:alp_lag}
    \mathcal{L} \supset \dfrac{1}{2}(\partial_\mu a)(\partial^\mu a) - \dfrac{1}{2} m_a^2 a^2 - i g_{ae} a \bar{e} \gamma^5 e - \dfrac{1}{4}g_{a\gamma}a F_{\mu\nu} \Tilde{F}^{\mu\nu}\,.
\end{equation}
Here, $a$ is the pseudoscalar ALP field with mass $m_a$, with coupling $g_{ae}$ to electrons and a dimension-5 coupling $g_{a\gamma}$ to photons. The electromagnetic field strength and its dual are denoted by $F_{\mu\nu}$ and $\Tilde{F}_{\mu\nu}$, respectively and $F_{\mu\nu} \Tilde{F}^{\mu\nu} \equiv F_{\mu\nu} F_{\alpha\beta} \epsilon^{\alpha\beta\mu\nu}$. These two couplings give rise to a rich assortment of scattering and decay channels, including the decays to electron-positron pairs $a \to e^+ e^-$ and photon pairs $a \to \gamma \gamma$. The variety of production and detection channels that arise from eq.~\eqref{eq:alp_lag} are presented in Table~\ref{tab:channels} and shown in Fig.~\ref{fig:feynman}. In our analysis, we will consider two scenarios where either the coupling to electrons ($g_{ae}$) or photons ($g_{a\gamma}$) is 
%dominant \newtext{while the other one is} comparatively smaller with a negligible contribution to the event rates. 
turned on while the other is taken to be zero. See Ref.~\cite{Liu:2023bby} for a discussion on the concurrent effects of both couplings. For this work we take the effects of these couplings at tree-level for the purposes of phenomenological simplicity and ease of comparison with other laboratory and astrophysical constraints in the literature. In principle, one may equivalently study derivative couplings to electrons instead, of the form $(\partial_\mu a) \bar{e} \gamma^\mu \gamma^5 e$, which can be approximately related to the Yukawa type operator $-i a \bar{e}\gamma^5 e$ via the Dirac equation in the weakly interacting limit. These two operators do exhibit different phenomenology, though, e.g. for multi-ALP interaction diagrams~\cite{Eberhart:2025lyu} and in long-range force effects~\cite{Ferrer:1998ue, Bauer:2023czj, Grossman:2025cov}. For the physics we consider in this work, discussed below, the Yukawa type and derivative type are approximately equivalent in their phenomenology.

\begin{table}[ht]
    \centering
    \begin{tabularx}{\textwidth}{lXX}
        \hline
         Coupling \, \, \, \,  & Production & Detection \\
         \hline
         $g_{a\gamma}$ & $\gamma \, A \to a \, A$ & $a \, A \to \gamma \, A$ \\
         & & $a \to \gamma \gamma$ \\
         \hline
         $g_{ae}$ & $\gamma \, e^- \to a \, e^-$ & $a \, e^- \to \gamma \, e^-$ \\
         & $e^+ \, e^- \to a \, \gamma$ & $a \to e^+ e^-$ \\
         & $e^\pm \, A \to e^\pm \, A \, a$ & $a \, A \to e^+ \, e^- \, A$  \\
         & $e^+ \, e^- \to a$ & \\
         \hline
    \end{tabularx}
    \caption{Considered ALP production and detection mechanisms through couplings to electrons and photons.}
    \label{tab:channels}
\end{table}

We simulate the generated flux of ALPs from all available channels by convolving the electron, positron, and photon fluxes simulated by \texttt{GEANT4} ~\cite{GEANT4:2002zbu, Allison:2006ve, Allison:2016lfl} in the DUNE graphite target (see \cref{sec:sims}) with the respective cross sections for each production channel in Table~\ref{tab:channels}. The simulation of the ALP flux from parent electron, positron, or photon \texttt{GEANT4} fluxes, and the subsequent transport and scattering/decay in the LAr detector, is performed with tools adapted from the \texttt{alplib} library~\cite{alplib}.

\begin{figure}
    \centering
    \includegraphics[width=0.8\textwidth]{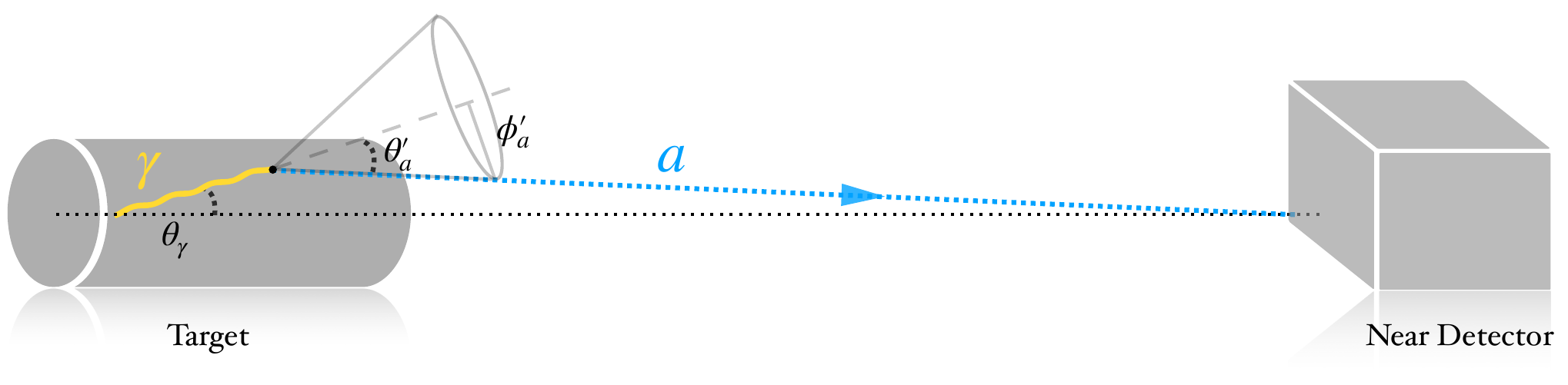}
    \caption{Schematic for the treatment of ALP flux, shown for a photon-produced ALP.
    ALP propagation to the detector is shown; the acceptance to the detector is decided by integrating the angular spectrum over the detector solid angle. Here, the angles $\theta_a^\prime, \phi_a^\prime$ are with respect to the parent photon direction, which can be translated to the laboratory coordinates via eq.~\eqref{eq:angle}.}
    \label{fig:dune-diagram}
\end{figure}

Once the ALPs have been produced, they should propagate 574 meters downstream of the target, where we account for their angular distribution such that only those that fall within the solid angle of the DUNE ND-LAr are able to scatter or decay within the detector fiducial volume. A schematic of the transport is shown in Fig.~\ref{fig:dune-diagram}, where the angle of the ALP in the laboratory frame ($\theta_a$) can be expressed in terms of the angle of the beam-induced photon/electron/positron parent ($\theta_\gamma$ in the figure), and the angles of the ALP production $\theta_a^\prime, \phi_a^\prime$ in the parent frame, as
\begin{equation}
    \theta_a = \arccos (\cos\theta_a^\prime \cos\theta_\gamma + \cos\phi_a^\prime \sin\theta_a^\prime \sin\theta_\gamma)\,.
    \label{eq:angle}
\end{equation}
For this solid angle, we take the DUNE ND-LAr as a rectangular prism with dimensions $3~{\rm m}\times7~{\rm m}$ in cross-sectional area and 5~m long, leading to an approximate solid angle acceptance of $\theta_a \lesssim \theta_{\rm det} \simeq 5$ mrad. This solid angle acceptance can be expressed with a Heaviside theta function in the flux integration that we discuss in what follows. 

\bigskip

\noindent {\textbf{ALP-photon coupling}}: In this case, the ALP production arises from Primakoff scattering of cascade photons inside the graphite target, $\gamma A \to a A$, where $A = ^{12}$C and the scattering takes place coherently with the combined atomic and nuclear electric fields. For the analytic form of the production cross-section, $\sigma^P$, and the atomic form factor  see refs.~\cite{Primakoff:1951iae,PhysRevLett.33.1400,Tsai:1986tx} and analyses in, e.g., refs.~\cite{Brdar:2020dpr,Brdar:2022vum,Jang:2022tsp,Capozzi:2023ffu,Dutta:2023abe,Kim:2024vxg}. For a parent photon rate in the target (in counts per unit energy per solid angle per protons-on-target (POT)), $d^2 N_\gamma / (dE_\gamma d\Omega_\gamma)$, the event rate of ALPs produced at the target through the Primakoff scattering is given as
\begin{align}
\label{eq:alp_prim_flux}
    \dfrac{dN_a^{P}}{dE_a} &=  \int \dfrac{1}{\sigma_\gamma (E_\gamma)}\frac{d^2 N_\gamma}{dE_\gamma d\Omega_\gamma}  \Theta(\theta_\text{det} - \theta_a) \dfrac{d^2\sigma_P}{dE_a d\Omega_a} \,d\Omega_a \,d\Omega_\gamma \, dE_\gamma \, ,
\end{align}
 where $\sigma_\gamma$ is the total photon absorption cross-section in the carbon target, $d\Omega_a = d\cos\theta_a^\prime d\phi_a^\prime$, and $d^2\sigma_P/(dE_a d\Omega_a)$ is the Primakoff differential scattering cross-section, which in the forward limit yields $E_a \simeq E_\gamma$. \\

Detection for ALPs with coupling to photons then takes place either through:
\begin{enumerate}
  \item ALP decays $a \to \gamma \gamma$.\\\\
   For this process, the decay rate reads
    \begin{equation}
        \Gamma(a\to\gamma\gamma) = \dfrac{g_{a\gamma}^2m_a^3}{64\pi} \, .
    \end{equation}
    The probability that ALP decays into two photons inside the detector is given by $P_{\rm decay}$, which is calculated by integrating over the probability density between the front and the back of the detector, $(\ell, \ell+\Delta\ell)$:
\begin{equation}
\label{eq:ProbDecay}
    P_{\rm decay}= e^{-\ell/(\tau v_a)} \left[ 1 - e^{-\Delta\ell /(\tau v_a) } \right]\,.
\end{equation}
Here, $\tau$ is the ALP lifetime in the laboratory frame,
$v_a = p_a / E_a$ is the ALP velocity, $\Delta\ell$ is the length of the DUNE ND-LAr over which the decay must happen (for which we take $\Delta\ell = 5$~m), and $\ell = 574$~m is the baseline distance from the source to the detector. For the total exposure in number of POT,  $\mathcal{E}$, the total ALP decay rate at the detector is given by a convolution of the decay probability with the ALP flux from eq.~\eqref{eq:alp_prim_flux}
\begin{equation}
\label{eq:ALPRateDecay}
    N^P_{\rm decay} = \mathcal{E}\int \frac{d N^P_a}{dE_a}P_{\rm decay} \,dE_a\,.
\end{equation}
This yields a $2\gamma$ signature in the detector, provided the two photons can be resolved through angular separation $\Delta\theta(\gamma\gamma) \gtrsim 1^\circ$~\cite{DUNE:2020ypp}. 

    \item Inverse Primakoff scattering $a A \to \gamma A$, producing a single photon final state from coherent scattering off liquid $^{40}$Ar.\\
    
    This process has a cross-section $\sigma_{IP}$ and is calculated in a similar manner to the production cross-section, but larger by a factor of 2 to account for the sum over polarization states. In this case, for a given exposure time $\mathcal{E}$, the number of ALPs scattering at the detector is given by
    \begin{equation}
    \label{eq:ALPRateScattering}
    N^{IP}_{\rm scatter} = \mathcal{E} \int n_T \ell \sigma_{IP}(E_a)\frac{d N^P_a}{dE_a} P_{\rm surv}\, dE_a\,,
    \end{equation}
    where $n_T$ is the number density of argon targets, $\ell$ is the detector length, and $P_{\rm surv}$ is the probability that an ALP survives to the detector without decaying, given by $P_{\rm surv}=e^{-\ell /(\tau v_a)}$.

\end{enumerate}

\noindent {\textbf{ALP-electron coupling}}: In this case, we have a richer set of production and detection channels at our disposal. ALPs can be sourced from the scattering of cascade electrons/positrons off $^{12}$C via bremsstrahlung, $e^\pm A \to e^\pm A a$, or positrons that annihilate atomic electrons via associated ($e^+ e^- \to \gamma a$) and resonant ($e^+ e^- \to a$) production. 
Cascade photons may also scatter off atomic electrons via the Compton-like process $\gamma e^- \to e^- a$. One should also consider charged meson decays $\pi^\pm \to e^\pm \nu_e a$ ($K^\pm \to e^\pm \nu_e a$), where the ALP $a$ is radiated off the electron current in a three-body decay, in the focusing-horn and decay volume complex~\cite{Dutta:2021cip,Dutta:2025fgz}. Starting instead with the ALP-electron couplings in the presence of the electroweak theory, running to the low-scale effective theory after electroweak symmetry breaking endows the ALP with additional couplings that can enhance the charged meson decays through contact interactions~\cite{PhysRevLett.130.241801}. In the minimal phenomenological setup adopted in this work, we estimate that the contribution from charged meson decays is largely subdominant, but a future dedicated analysis with a simulation of the meson focusing and the aforementioned electroweak vertices included is highly motivating.

For Compton scattering production of ALPs, the event rate of ALPs produced at the target pointing within the solid angle of the detector is
\begin{align}
\label{eq:alp_comp_flux}
    \dfrac{dN_a^{\rm C}}{dE_a} &= \int \dfrac{1}{\sigma_\gamma (E_\gamma)} \frac{d^2 N_\gamma}{dE_\gamma d\Omega_\gamma}  \Theta(\theta_\text{det} - \theta_a) \dfrac{d^2\sigma^{\rm C}(E_\gamma)}{dE_a d\Omega_a}\,dE_\gamma\,d\Omega_\gamma  \,d\Omega_a  \,,
\end{align}
where ${d^2\sigma^{\rm C}(E_\gamma)}/{dE_a d\Omega_a}$ is the differential Compton cross-section with respect to the ALP energy and solid angle. The analytical formula for $\sigma^{\rm C}$ can be found in refs.~\cite{Tsai:1986tx,Fukugita:1982ep,Gondolo:2008dd}.

For electron-induced or positron-induced ALP fluxes, additionally, we must account for the energy loss of the parent electron/positron in material by convolution with the track length distribution function for $e^\pm$ radiative transport, $I(t, E_e, E_e^\prime)$, for an electron/positron initial ($E_e$) and final ($E_e^\prime$) state energy and distance traveled $t$, the dimensionless number of radiation lengths.
For the carbon-based target, we take one radiation length as $42.7$ g/cm$^2$. The flux prediction for ALPs whose momenta point within the detector solid angle can then be written as
\begin{align}
\label{eq:alp_epem_flux}
    \dfrac{dN^r_a}{dE_a} &= \frac{N_A X_0}{A} \int \frac{d^2 N_{e}}{dE_e d\Omega_e}  I(t, E_e, E_e^\prime)  \Theta(\theta_\text{det} - \theta_a) \dfrac{d^2\sigma^{r}(E^\prime_e)}{dE_a d\Omega_a} d\Omega_e \,d\Omega_a \,dE_e\, dt\, dE^\prime_e \, ,
\end{align}
 where $N_A$ is Avogadro's number, $X_0$ is the radiation length in g/cm$^2$, $A$ is the molar mass of carbon in g/mol, $d^2\sigma^r / dE_a d\Omega_a$ is the differential cross-section with respect to the ALP energy and solid angle, for $r$ being either the resonance production, the associated production or bremsstrahlung. The analytical formulae for each relevant cross-section can be found in, for example, refs.~\cite{CCM:2021jmk,AristizabalSierra:2020rom,PhysRevD.34.1326,Avignone:1988bv,Gondolo:2008dd}. In the above integral, we integrate over the electron/positron flux $d^2 N_{e}/{dE_e d\Omega_e}$ (number per unit energy per solid angle per POT).

Energy loss from transport in carbon leads to a final energy $E^\prime_e$ after $t$ radiation lengths, where ALP creation takes place via the differential cross-section (bremsstrahlung, associated production, or resonant production). In this way, ALP with energy $E_a$ and solid angle $\Omega := (\theta_a, \phi_a)$ is produced.  Here, the analytical approximation for the energy loss function is~\cite{PhysRevD.34.1326}
\begin{equation}
    I(t, E_e, E_e^\prime) = \frac{\Theta(E_e - E_e^\prime)}{E_e \Gamma (4 t/3)} (\ln E_e/E_e^\prime)^{4t/3 - 1} \, .
\end{equation}
Note that integration over the dimensionless track length $t$ runs from $0$ to $T$, where $T = \rho L / X_0$. Here, $L$ is a path length through the target (cm) and  $\rho$ (g/cm$^3$) is the mass density. This integration over $dt$ in eq.~\eqref{eq:alp_epem_flux} can be done explicitly by numerical integration, as $\int_0^T I(t,E_e,E_e^\prime) dt$ does not have a known closed form expression. 

Note that modeling electron/positron energy loss for bremsstrahlung and resonant production using the track-length distribution function does not include any dependence on the angular distribution of the electron/positron after traversing $T$ track lengths. In other words, the random walk behavior of the electromagnetic shower particles is not captured. In ref.~\cite{Blinov:2024pza}, the \texttt{PETITE} package demonstrated that explicitly modeling the shower production of weakly interacting vector particles can lead to significantly different flux estimates due to the aforementioned angular spreading of the parent electron/positron flux. A similar result should hold for pseudoscalar production, as in our case. To account for the broadening of the ALP flux, we limit the track length integral up to $T = 5$ radiation lengths in the ALP flux integral given in eq.~\eqref{eq:alp_epem_flux}. We find that our results agree with the \texttt{PETITE}-derived fluxes within a 20\% margin.\\

The detection of ALPs with couplings to electrons can then take place via several channels:
\begin{enumerate}
    \item ALP decays to $e^+ e^-$ pairs.\\\\
    For this process, the decay width reads 
\begin{equation}
    \Gamma(a \to e^+ e^-) = \frac{g_{ae}^2 m_a}{8 \pi} \sqrt{1 - \frac{4 m_e^2}{m_a^2}} \, ,
\end{equation}
 similarly to ALP decay to two photons. For the total exposure in POT, $\mathcal{E}$, the total ALP decay rate at the detector is given by
\begin{equation}
    N^C_{\rm decay} = \mathcal{E}\int \sum_i \frac{d N_a^i}{dE_a}P_{\rm decay} \,dE_a\,,
\end{equation}
where ${d N^i_a}/{dE_a}$ are the ALP fluxes of type $i=\,$Compton, bremsstrahlung, associated, and resonant production outlined in the above discussion, and the decay probability is the same as in eq.~\eqref{eq:ProbDecay}.
\item Scattering via $e^+ e^-$ pair production.\\\\
This is a process similar to Bethe-Heitler pair production by photons ($a A \to e^+ e^- A$), with cross-section $\sigma^{e^+e^-}$. The calculational details for this cross-section are described in Appendix.~\ref{app:pair_prod}. This results in a total event rate at the detector given by 
\begin{equation}
    N^{e^+e^-}_{\rm scatter} = \,\mathcal{E} \int \sum_i n_T^e \,  \ell \, \sigma^{e^+e^-}(E_a)\frac{d N^i_a}{dE_a} P_{\rm surv}\, dE_a\,,
    \end{equation}
where $n_T^e$ is the number density of electron targets in the detector and $\ell$ is the detector length.
This process leaves a signature of a very forward $e^+ e^-$ pair with small angular separation between the two leptons after scattering coherently via photon exchange with the atomic target. 
\item Inverse Compton scattering ($a e^- \to \gamma e^-$).\\

 The total number of scattering events at the detector is given by
\begin{equation}
    N^{IC}_{\rm scatter} =  \mathcal{E} \int \sum_i n_T^e \, \ell \, \sigma^{IC}(E_a)\frac{d N^i_a}{dE_a} P_{\rm surv}\, dE_a\,,
    \end{equation}
where $\sigma^{IC}$ is the cross-section for inverse Compton scattering~\cite{Avignone:1988bv}.

\end{enumerate}

In addition, if one takes this phenomenological model beyond leading order, electron loop diagrams induce an effective coupling to electroweak bosons including photons. In principle, this would also induce $a \to \gamma\gamma$ through this induced coupling, importantly opening a secondary decay channel below $m_a < 2 m_e$. This possibility has been considered, for example, in ref.~\cite{Bauer:2020jbp} and exhibits new bounds from beam dump experiments that account for $\gamma\gamma$ decays~\cite{Liu:2017htz}. Again, for this work we shall omit the contributions of this channel below $m_a = 2 m_e$ for the purposes of easier comparison with laboratory and astrophysical constraints such as NA62 and stellar cooling bounds that appear below this mass limit.

%%%%%%%%%%%%%%%%%%%%%%%%%%%%%%%%%%%%%%%%%%%%%%%%%%%%%%%%%%%%%%%%%%%%%%%%%%%%%%

%%%%%%%%%%%%%%%%%%%%%%%%%%%%%%%%%%%%%%%%%%%%%%%%%%%%%%%%%%%%%%%%%%%%%%%%%%%%%%
\section{\texttt{GEANT4} and \texttt{GENIE} Simulation}\label{sec:sims}
In the DUNE experiment, a 120 GeV proton beam hits a carbon target, producing particles of interest such as electrons and photons. High-energy photons are primarily generated through intranuclear interactions during proton-carbon collisions where various mesons, especially neutral mesons (like pions and eta), decay and emit pairs of photons. Electrons are produced through processes such as bremsstrahlung and pair production, which are relevant to developing electromagnetic showers, and such processes can be effectively simulated by the \texttt{GEANT4} ~\cite{GEANT4:2002zbu, Allison:2006ve, Allison:2016lfl} toolkit. In order to model the fluxes of $e^+$,  $e^-$, and $\gamma$ inside the target, we simulate cylindrical carbon targets with the same configuration as it was described in DUNE Technical Design Report \cite{DUNE:2020ypp} and we have used \texttt{QGSP\_BERT} physics list for the hadronic processes and \texttt{G4EmStandardPhysics} for the electromagnetic reactions. To record kinematic information of particles produced in the target, we implemented user-defined class of \texttt{G4UserSteppingAction} derived from an abstract class \texttt{G4SteppingAction}.

\subsection{Signal Fluxes from the Target}
\begin{figure}[ht!]
    \centering
    \includegraphics[width=1.0\textwidth]{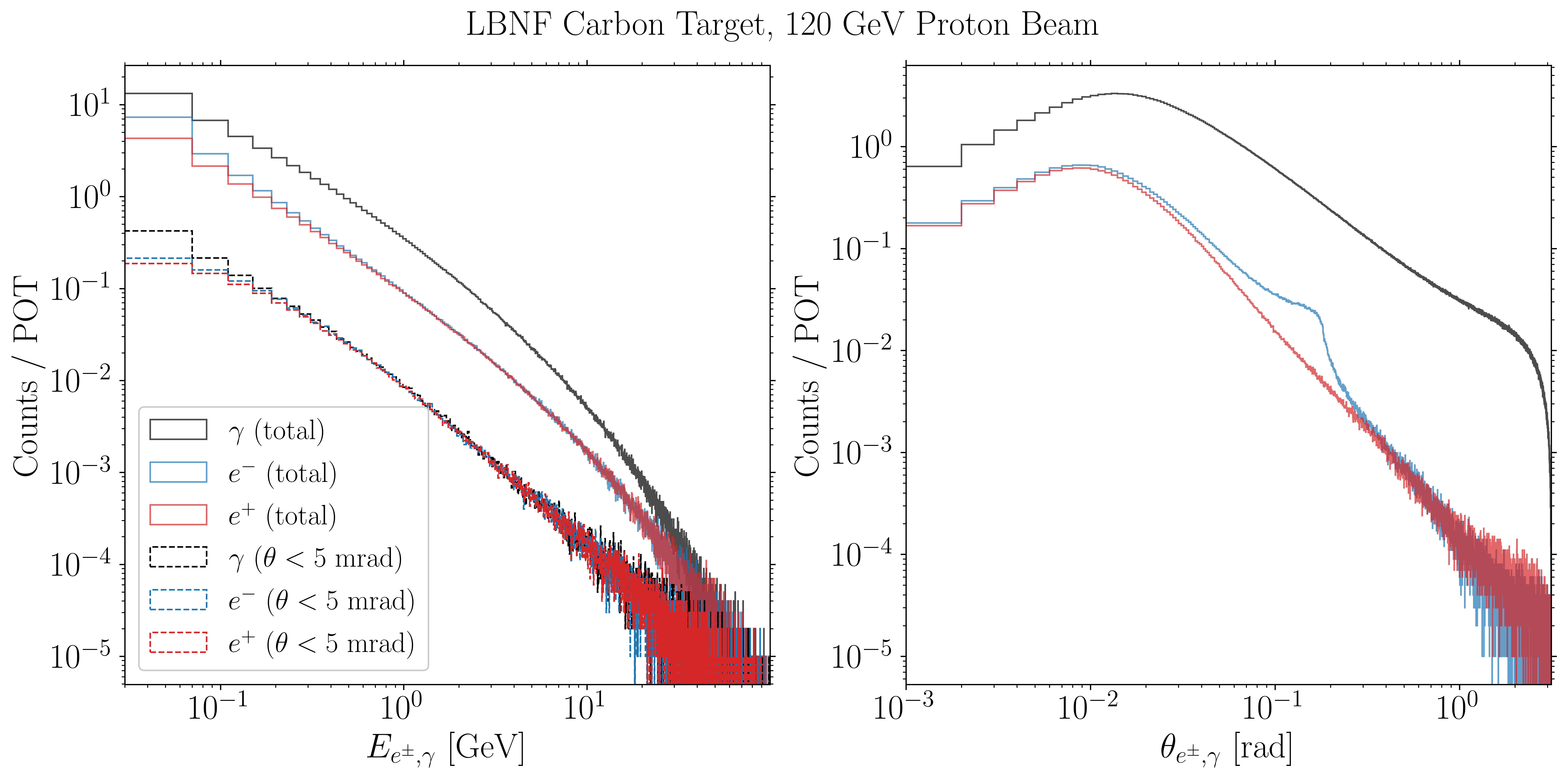}
    \caption{One-dimensional flux histograms: Photon, positron, and electron fluxes in the DUNE graphite target from \texttt{GEANT4} simulation based on a $10^5$ POT sample, binned in steps of 20 MeV over the $e^\pm /\gamma$ energies (left) and their angles with respect to the beam axis binned in steps of 1 mrad (right) at production. In the left panel, we also show the energy distributions of the forward component of the fluxes after a 5 mrad angular cut (dotted histograms). The shallow ``knee'' over 0.1-0.2 rad region in the electron angular spectrum in the right panel can be interpreted as a result of Compton scattering by high-energy cascade electrons and $\gamma$-rays.}
    \label{fig:dune_fluxes_1d}
% \end{figure}
\vspace{0.5cm}
% \begin{figure}
%     \centering
    \includegraphics[width=1.0\textwidth]{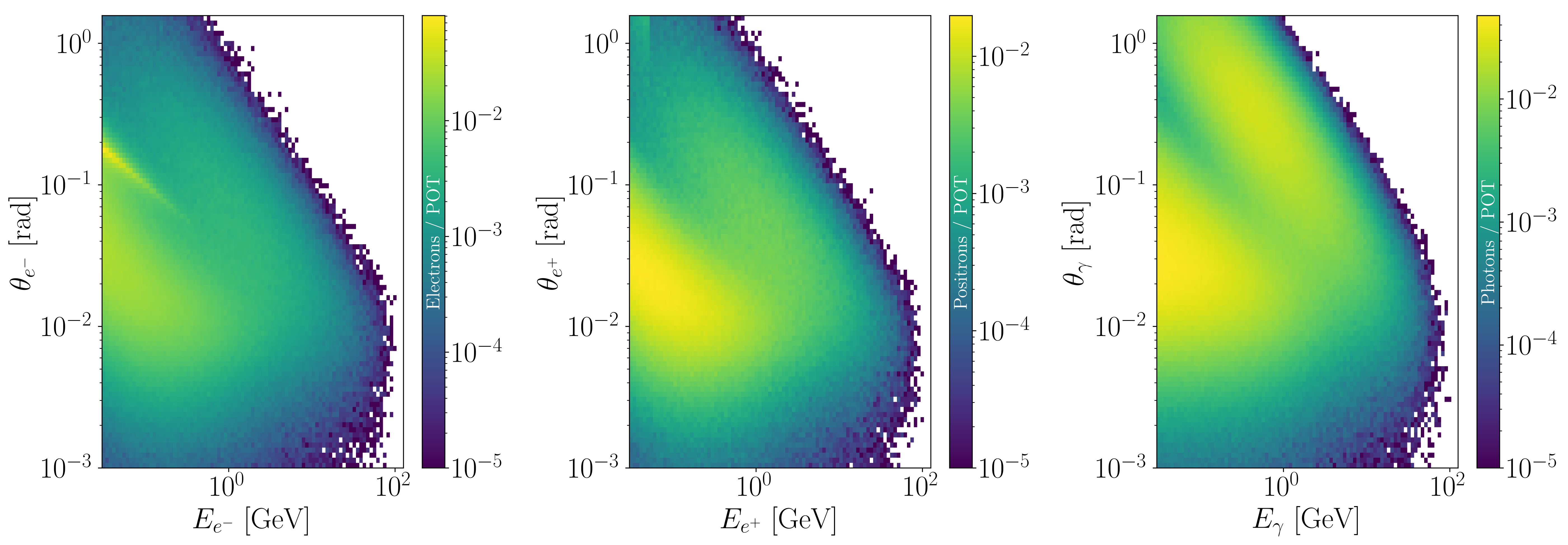}
    \caption{Two-dimensional flux histograms: Photon, positron, and electron fluxes in the DUNE graphite target from \texttt{GEANT4} simulation based on a $10^5$ POT sample, binned over the $e^\pm /\gamma$ energies and their angles at production with respect to the beam axis.}
    \label{fig:dune_fluxes_2d}
\end{figure}

In Fig.~\ref{fig:dune_fluxes_1d}, the fluxes of selected electrons, positrons, and photons produced in the DUNE target, relevant for the production channels listed in Table~\ref{tab:channels}, are shown as 1-dimensional histograms over their kinetic energies and angles with respect to the beam axis. The shallow ``knee'' observed in the electron angular distribution around $\theta$ in 0.1-0.2 rad is best explained by high-energy electrons produced via Compton scattering of cascade electrons and photons produced inside the target material. In Fig.~\ref{fig:dune_fluxes_2d} we show fluxes of electrons, positrons and photons in the form of 2-dimensional histograms. These fluxes are then convolved with the respective cross sections for the ALP production mechanisms in \cref{tab:channels} as we have explained in the previous section. Since the output from \texttt{GEANT4} is a Monte Carlo sample of discrete electrons, positrons, and photons, in ALP flux calculations we perform a sum instead of the integration; for example, the integration over electron energies and angles is replaced by
\begin{equation*}
    \int dE_e d\Omega_e \to \sum_{E_{e,i}, \theta_{e,i}} N_i \,,
\end{equation*}
where $N_i$ is a histogram weight carrying the number of electrons per POT with energy $E_{e,i}$ and $\theta_{e,i}$ is an angle with respect to the beam axis. We integrate uniformly over $d\phi_e$ due to the azimuthal symmetry of the target.

\subsection{The Escaped Proton Fraction: Signal Fluxes from the Beam Dump}
In addition to the fluxes of secondary particles produced from protons impinging on the DUNE graphite target, there is a non-negligible chance that some protons will escape the target without getting absorbed. These protons are then free to traverse the decay pipe until impinging again on the dump in the absorber hall, roughly 270 meters downstream of the target. The dump itself can be described roughly as a $4\times 4 \times 4$ m$^3$ aluminum cube whose center is 306 m from the center of the DUNE near detector.

From \texttt{GEANT4} simulation of the proton transport through the graphite target, we find that approximately 4.25\% of protons escape the target and propagate through the decay pipe (through air with modest attenuation) to reach the dump. At the face of the dump, the cross-sectional $(x,y)$ profile of the proton bunch is broader compared to its prompt on-target profile, but we find that the resulting $(x,y)$ distributions of secondary particles ($e^\pm$, $\gamma$) are still tightly localized in $(x,y)$ to a radius of around 10 cm from the center of the beamline. Overall, the kinematic distributions of the secondary photons, electrons, and positrons in the LBNF beam dump are very similar to those of the target shown in Fig.~\ref{fig:dune_fluxes_1d}, only reduced in normalization per POT.

\subsection{Neutrino Detector Background Simulation}
Electrons, positrons, and gamma rays produced from neutrino interactions in the detector act as backgrounds for our BSM study, and their production processes can be estimated through simulations. Using \texttt{\texttt{GENIE}-MC}~\cite{Andreopoulos:2015wxa}, we input the DUNE neutrino flux profile~\cite{DUNE:2020ypp} into the simulation. In this simulation, we have used the default tuning configuration of \texttt{\texttt{GENIE}-MC 3.02.00}. The stable final state particles generated by neutrino interactions within the liquid argon volume are recorded for analysis. In the process of data analysis, we made preselection against collected final-state particles, and we also performed detailed selection criteria; both are discussed in \cref{sec:selection}.

Finally, for both the signal and background modeling of final state photons, electrons, and positrons, we account for the smearing of true energies into reconstructed energies through the use of an energy resolution function. In principle, the calorimetry of photons and electrons/positrons may differ, but for simplicity, we adopt a singular resolution function for all electromagnetic showers. Given a true energy $E_{truth}$ of the final state photon, electron, or positron generated by Monte Carlo for either the background or the signal, we adopt a normal distribution centered at $E_{truth}$ whose width is given by
\begin{equation}
    \frac{\sigma(E_{truth})}{E_{truth}} =  a + \frac{b}{\sqrt{E_{truth}}} + \frac{c}{E_{truth}}  \, ,
\end{equation}
where the coefficients $a$, $b$, and $c$ are adopted from ref.~\cite{Rout:2025jdi}; $\{a,b,c\}=\{0.027, 0.024, 0.007\}$. This energy scaling leads to energy resolutions in the 15\% - 5\% range at energies below and above 1 GeV, respectively; see also ref.~\cite{Marshall:2019vdy} for electron-specific considerations. For other work on DUNE's potential calorimetry capabilities, see refs.~\cite{DUNE:2021cuw,Friedland:2018vry,DeRomeri:2016qwo,Chatterjee:2021wac}.

%%%%%%%%%%%%%%%%%%%%%%%%%%%%%%%%%%%%%%%%%%%%%%%%%%%%%%%%%%%%%%%%%%%%%%%%%%%%%%%%%%
\section{Preselection Cuts and Final State Topologies}\label{sec:selection}
The ALP signal hypothesis has several possible final states from scattering or decays in the detector: $e^+ e^-$ (decays and pair production), $2\gamma$ (decays), $1\gamma$ (inverse Primakoff), and $1\gamma1e^-$ (inverse Compton scattering). Each of these final states should have no other electromagnetic or hadronic activity, so the particle identification and rejection of extraneous particles will be necessary.

In Table~\ref{tab:cuts} we list the kinetic energy thresholds and angular resolutions of identifiable particles produced in the LArTPC. As a post-processing step after neutrino-induced background simulation with \texttt{GENIE}, we check each event against these cuts and place it into one of the four topologies we have considered for this analysis. Events containing particles not contributing to the final states of interest, with their kinetic energies exceeding the respective threshold values listed in Table~\ref{tab:cuts}, are discarded. Therefore, at the truth level, an accepted event may have, for example, soft hadronic or electromagnetic activity that lies below threshold, while only the hard activity in the event satisfies the topology criterion. We henceforth call these kinematic cuts on additional activity in the event \textit{preselection cuts}.

\begin{table}[ht]
    \centering
    \begin{tabular}{|c|c|c|}
    \hline
         Particle Type & Kinetic Energy Threshold &  Angular Resolution \\
         \hline
         $\mu^\pm$ & 30 MeV &  1$^\circ$\\
         $\pi^\pm$ & 100 MeV& 1$^\circ$ \\
         $e^\pm / \gamma$& 30 MeV &  1$^\circ$\\
         Protons & 50 MeV & 5$^\circ$ \\
         Neutrons & 50 MeV&  5$^\circ$\\
        other & 50 MeV & 5$^\circ$\\
         \hline
    \end{tabular}
    \caption{Energy threshold and angular resolution for different particle types. Taken from the DUNE TDR~\cite{DUNE:2020ypp}.}
    \label{tab:cuts}
\end{table}

Additionally, we pay special care to events with two-particle final states for $2\gamma$ and $e^+ e^-$; if their angular separation is less than 1$^\circ$, they would instead be reconstructed as a single electromagnetic shower, most likely appearing as a single photon or a single $e^\pm$ in the LArTPC. This applies to the signal hypothesis as well. We treat these cases as mis-identification (mis-ID) events that contaminate the single-photon/electron/positron final state. For tagging and finding the mis-ID rates, we have used the $dE/dx$ study in ArgoNeuT~\cite{ArgoNeuT:2016wjb}, the MicroBooNE study of using convolutional neural networks in ref.~\cite{MicroBooNE:2016dpb}, the $1\gamma0p$ and $2\gamma0p$ studies\footnote{The notation here signifies one or two photons with zero hadronic activity in the final state.} of ref.~\cite{MicroBooNEPRL}, and finally the information from Fig.~1.4 of the DUNE TDR~\cite{DUNE:2020ypp}. Given all that, we obtained the conservative estimate for $e^\pm \leftrightarrow \gamma$ mis-ID rate of 18\%. We also do not assume any charge identification capability in the LAr module, treating $e^+$ and $e^-$ final states as identical in this analysis; therefore, the $e^+ e^-$ and $1\gamma 1e^-$ ALP signals are effectively treated as $e^\pm e^\pm$ and $1\gamma 1e^\pm$ final states, respectively.

The mis-ID probability may affect cross-contamination of $e^+ e^-$, $1\gamma1e^\pm$, $1\gamma$, and $2\gamma$ final states; for example, if a true $e^\pm e^\pm$ final state (post preselection) has one of its electrons/positrons mis-identified as a photon, thereby being reconstructed as a $1\gamma 1e^\pm$ final state. This chance goes according to the binomial probability of having one ``success'' (mis-ID) for two ``trials'' (particles) at a $18\%$ rate;
\begin{equation}
    P(e^\pm e^\pm \to 1\gamma 1e^\pm) = f_{\rm binom} (1,2,0.18) = 29.52\%\,.
\end{equation}
Likewise, the $2\gamma$ final state may get mis-ID'd into the $1\gamma 1e^\pm$ final state with the same probability. The $1\gamma 1e^\pm$ final state can be mis-ID'd into either $e^\pm e^\pm$ or $2\gamma$ at the same binomial rate, contributing half ($14.76\%$) to each channel. The reduction of these two-particle final states by 29.52\% and their growth from cross-contaminating mis-ID'd channels is reflected in the total rates.

Finally, we also account for the contribution to backgrounds due to the loss of containment of photons that shower outside of the fiducial volume of the detector. This is performed as a post-processing step after the \texttt{GENIE} simulation. We find that around $1\%$ of $2\gamma$ events will result in one of the photons pair-converting outside the detector fiducial volume, consistent with ref.~\cite{Coloma:2023oxx}. However, the remaining photon in the fiducial volume further tends to pair-convert to a very collinear $e^+ e^-$ pair via Bethe-Heitler scattering with less than $1^\circ$ in the opening angle; therefore, we treat these events as contributions to the single shower, $1\gamma$ final state background, rather than the $e^+ e^-$ background rate in the labeling scheme of ref.~\cite{Coloma:2023oxx}.

We show the number of neutrino-induced background events for each topology in \cref{tab:acceptances}, assuming 3.5 years of running in the Forward Horn Current (FHC) neutrino mode. The numbers are based on the \texttt{GENIE} output of our background simulation and broken down by source neutrino flavor ($\nu_\mu$, $\nu_e$, $\bar{\nu}_\mu$, and $\bar{\nu}_e$). Each block row separates the background events by final state, and the subrows distinguish the event rates due to a $\gamma \leftrightarrow e^\pm$ mis-identification rate of 18\%. Since the \texttt{GENIE} simulation was computed for $10^7$ neutrino interactions, we report the event rates in \cref{tab:acceptances} by counting the events in each final state per parent neutrino flavor (after our preselection criteria discussed above), then multiply by the total number of expected neutrino events in 3.5 years ($5.145 \times 10^{21}$ POT) for that flavor, and finally divide by $10^7$. 

As a consistency check, we compare these background event rates to those reported in ref.~\cite{Coloma:2023oxx} for the $e^+ e^-$ and $2 \gamma$ final states. Firstly, a true $e^+ e^-$ rate originates from \texttt{GENIE} simulation of the somewhat rare ``Dalitz'' decay $\pi^0 \to \gamma e^+ e^-$ (branching ratio of around 1.2\%~\cite{ParticleDataGroup:2024cfk}), in which the final state photon's energy is below the reconstruction threshold, resulting in a sole $e^+ e^-$ pair after preselection. The rate of $e^+ e^-$ reconstructed this way in our simulation is relatively small, amounting to less than 100 events per 3.5 years of exposure. On the other hand, ref.~\cite{Coloma:2023oxx} considered contributions to $e^+ e^-$ from $\pi^0 \to \gamma\gamma$ decays after one of the photons pair converts outside the fiducial volume, leaving only one photon to pair convert to $e^+ e^-$ inside the detector via Bethe-Heitler scattering. As mentioned above, our simulations of the loss due to containment result in a similar rate of single photons left within the fiducial volume (around 1\% of the total $2\gamma$ events, or 7663 single photons per 3.5 years after a mis-ID loss of $1 - 0.18 = 0.82$ is applied).
However, these photon pairs convert to very collinear $e^+ e^-$ pairs ($\ll 1^\circ$ separated) and we therefore instead treat them like single-shower, ``$1\gamma$'' events.

Comparing with ref.~\cite{Coloma:2023oxx} in the $2\gamma$ channel backgrounds, we find a few times larger number of events. This difference partially stems from the assumed reconstruction efficiency, which was inspired by MicroBooNE analysis of electromagnetic reconstruction in LAr~\cite{PhysRevD.107.012004}, adopted for DUNE in ref.~\cite{Coloma:2023oxx}; our analysis explicitly includes preselection and topological cuts that account for this efficiency (but not efficiencies of more bespoke selection cuts such as boosted decision trees as in ref.~\cite{PhysRevD.107.012004}). Our analysis is also performed explicitly at the DUNE energy scale, to which efficiencies in the context of the 8 GeV Booster proton beam may not scale appropriately. The present work and ref.~\cite{Coloma:2023oxx} do agree on the rate for neutral current $\pi^0$ production from \texttt{GENIE} output, a major contributor to the $2\gamma$ backgrounds, while the \texttt{GENIE} output in this work also accounts for charged current $\pi^0$ production, around 10 times larger than the neutral current rate.

\begin{table}[h]
    \centering
\resizebox{\textwidth}{!}{
    \begin{tabularx}{\textwidth}{|l|R|R|R|R|R|}
    \hline
    { Signal Topology} & \Centering $\nu_\mu$ & \Centering $\nu_e$  & \Centering $\bar{\nu}_\mu$  & \Centering $\bar{\nu}_e$  & \Centering Total Counts  \\
         \hline
         CC + NC $\nu$ events & $3.49\times10^8$ & $4.78\times10^6$ & $1.72\times10^7$ & $6.59\times 10^5$ & $3.71\times10^8$\\
         \hline
         \hline
         $\mathbf{e^+ e^-}$ & 49 & 1 & 8 & 0 & 58  \\
         $e^- (\gamma)^*$ & 123 & 76 & 9 & 0 & 208 \\
         $e^+ (\gamma)^*$  & 139 & 2 & 10 & 11 & 162 \\
         \hline
         $\mathbf{2\gamma}$ & 372,508 & 5,450 & 38,236  & 1,433 & 417,628 \\
         $1(e^-)^* 1\gamma$  & 123 & 76 & 9 & 0 & 208   \\
         $1(e^+)^* 1\gamma$  & 139 & 2 & 10 & 11 & 162  \\
         \hline
         $\mathbf{1e^- 1\gamma}$  & 590 & 363 & 41 & 2 & 996 \\
         1$e^+ \gamma$  & 663 & 7 & 47 & 51 & 768  \\
         $1e^\pm 1(e^\pm)^* $  & 21 & 0 & 4 & 0 & 25  \\
         $\gamma (\gamma)^* $  & 156,022 & 2,283 & 16,015 & 600 & 174,920 \\
         \hline
         $\mathbf{1\gamma}$ & 34,469 & 441 & 3,094 & 101 &  38,104 \\
         $1(e^-)^*$  & 3,839 & 6,516 & 417 & 46 & 10,818 \\
         $1(e^+)^*$  & 0 & 0 & 0 & 1,451 & 1,451 \\
         $1\gamma (1\gamma)_{\rm c}$ & 7,030 & 78 & 537 & 17 & 7,663 \\
         \hline
    \end{tabularx}
}
    \caption{The numbers of neutrino-induced background events (rounded up to the nearest integer) after preselection cuts in each signal topology, separated by each incoming neutrino flavor. The rates are based on 3.5 years of running in FHC mode and a simulation sample of $10^7$ neutrino interactions per flavor flux. The starred (*) particles involve a mis-identification (mis-ID) of $\gamma \leftrightarrow e^\pm$ that would fake the primary topologies (bold font) in each row group. Parentheses with ``c'' subscript indicate contributions from topologies in which the parenthesized particle is lost due to containment outside the fiducial detector volume. For comparison, the CC neutrino events shown in the first line are in good agreement with the rates presented at Table 6.2 of the DUNE ND CDR~\cite{DUNE:2021tad}.}
    \label{tab:acceptances}
\end{table}

%%%%%%%%%%%

\subsection{$\gamma\gamma$ Final State}
\label{sec:2gamma_fs}
 ALPs which couple to photons via $g_{a\gamma}$ may decay into di-photon pairs ($\gamma\gamma$). If the photon pair is not too collinear or soft ($\Delta\theta_{\gamma\gamma} > 1^\circ$ and $E_\gamma > 30$ MeV as per Table~\ref{tab:cuts}), then each photon is treated as separately resolved. For ALP masses below $\sim 500$ keV, nearly 100\% of all photon pairs from decays cannot be resolved due to the large relativistic boost the ALP would receive simply by virtue of the large separation between the typical momentum scale of the progenitor photons and the ALP mass. Above this mass, smaller couplings $g_{a\gamma}$ will require less boost in order for the ALP to live long enough to decay in the detector, so nearly all of the decaying ALPs above $m_a \sim 500$ keV will be resolved as di-photon pairs.
 
\begin{figure}[t]
    \centering
    \includegraphics[width=1.0\textwidth]{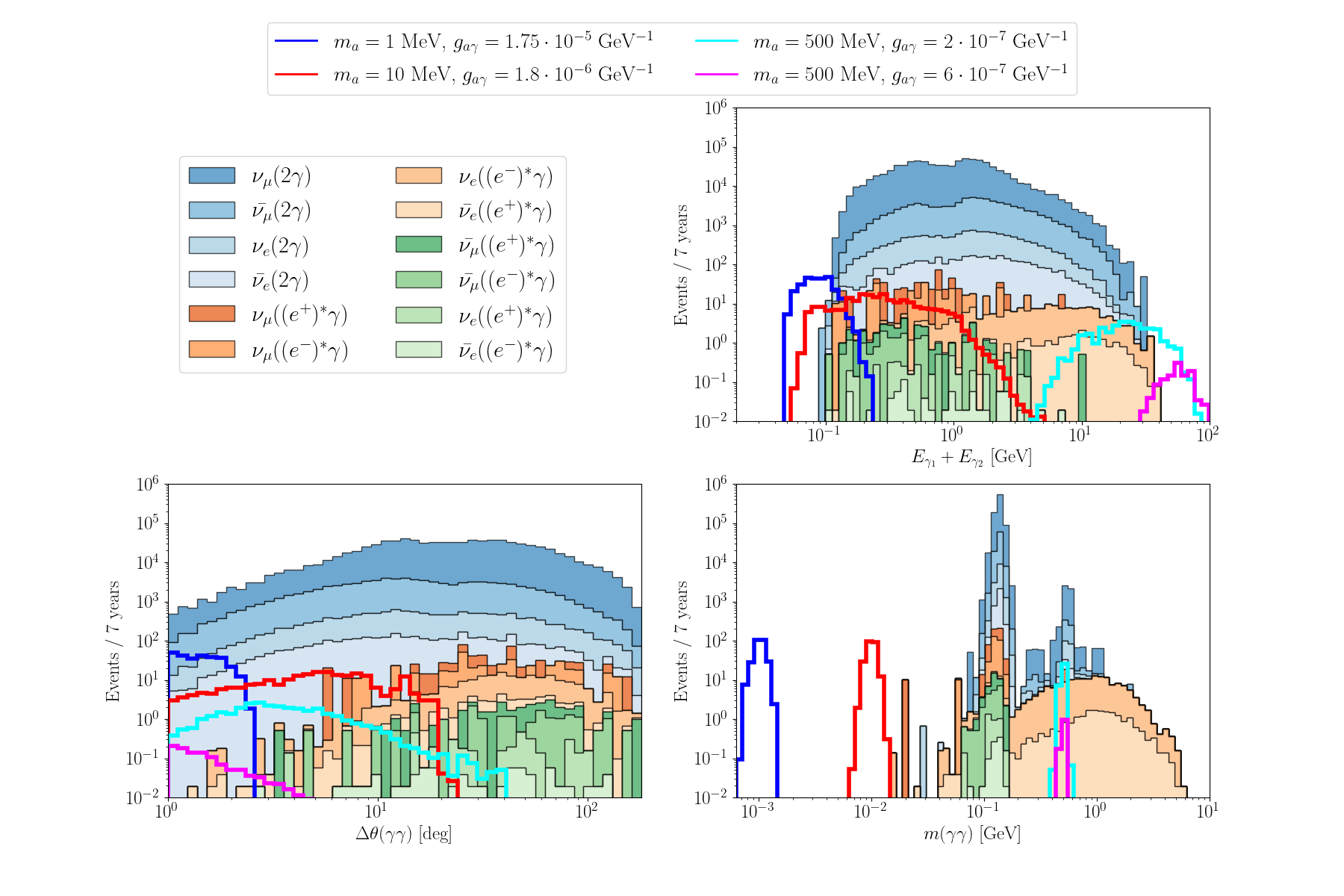}
    \caption{Di-photon final state: we show the total energy spectrum (upper right), opening angle (bottom left), and invariant mass (bottom right) distributions for the neutrino-induced backgrounds and the ALP signal. The ALP mass and coupling benchmark values are chosen to yield statistically significant event rates in the limit of smaller $g_{a\gamma}$ couplings (cyan and pink histograms) or larger coupling (dark blue and red histograms). The neutrino-induced backgrounds include the rates from $e^+$ and $e^-$ mis-ID that contaminate the $2\gamma$ signal as $(1e^+)^* \gamma$ or $(1e^-)^* \gamma$.}
    \label{fig:2g_kinematics}
\end{figure}

In Fig.~\ref{fig:2g_kinematics}, we show the di-photon distributions as a function of the total true energy  (top right), the opening angle $\Delta\theta(\gamma,\gamma)$ (bottom left), and the invariant mass $m(\gamma,\gamma)$ (bottom right), which is defined as
\begin{equation}
    m^2(\gamma,\gamma) = (p_{\gamma_1} + p_{\gamma_2})_\mu(p_{\gamma_1} + p_{\gamma_2})^\mu\,.
\end{equation}
For comparison, we also show the signal for ALPs of masses $m_a=1, 10$, and 500 MeV. For each of these benchmark points, we take different values of $g_{a\gamma}$ indicated in the legend. The smaller couplings, which will lie on the lower edge of the sensitivity limit for DUNE ND-LAr (shown later in \cref{sec:sensitivity}),  give rise to characteristically long-lived ALPs that require less boost from their progenitor photons in the target in order to reach the ND without prematurely decaying. Hence, they have softer energy spectra. Compare, for example, the $m_a = 500$ MeV mass point at $g_{a\gamma} = 2 \cdot 10^{-7}$ GeV$^{-1}$ (light blue) to that of $g_{a\gamma} = 6 \cdot 10^{-7}$ GeV$^{-1}$ (pink) in the histogram in \cref{fig:2g_kinematics}; for the larger coupling, the lifetime is shorter and therefore only the ALPs with large boost factor will decay in the detector. The ALP energy spectrum can either be very soft or very hard depending on the coupling due to the required boost factor, while the opening angle between the two photons remains relatively small compared to the neutrino backgrounds. Additionally, even after energy resolution effects, the invariant mass of the ALP remains relatively well-resolved in the background-dominated region. The findings shown in the lower panels of \cref{fig:2g_kinematics} motivate an opening angle cut of $\Delta\theta(\gamma,\gamma) < 20^\circ$ and an invariant mass cut of $|m(\gamma,\gamma) - m_a| < 0.05 \times m_a$ which can be checked to ensure more than 90\% signal efficiency in the majority of the ALP parameter space.

\subsection{Single Photon Final State}
\label{sec:single_gamma_fs}
\begin{figure}[t]
    \centering
    \includegraphics[width=1.0\textwidth]{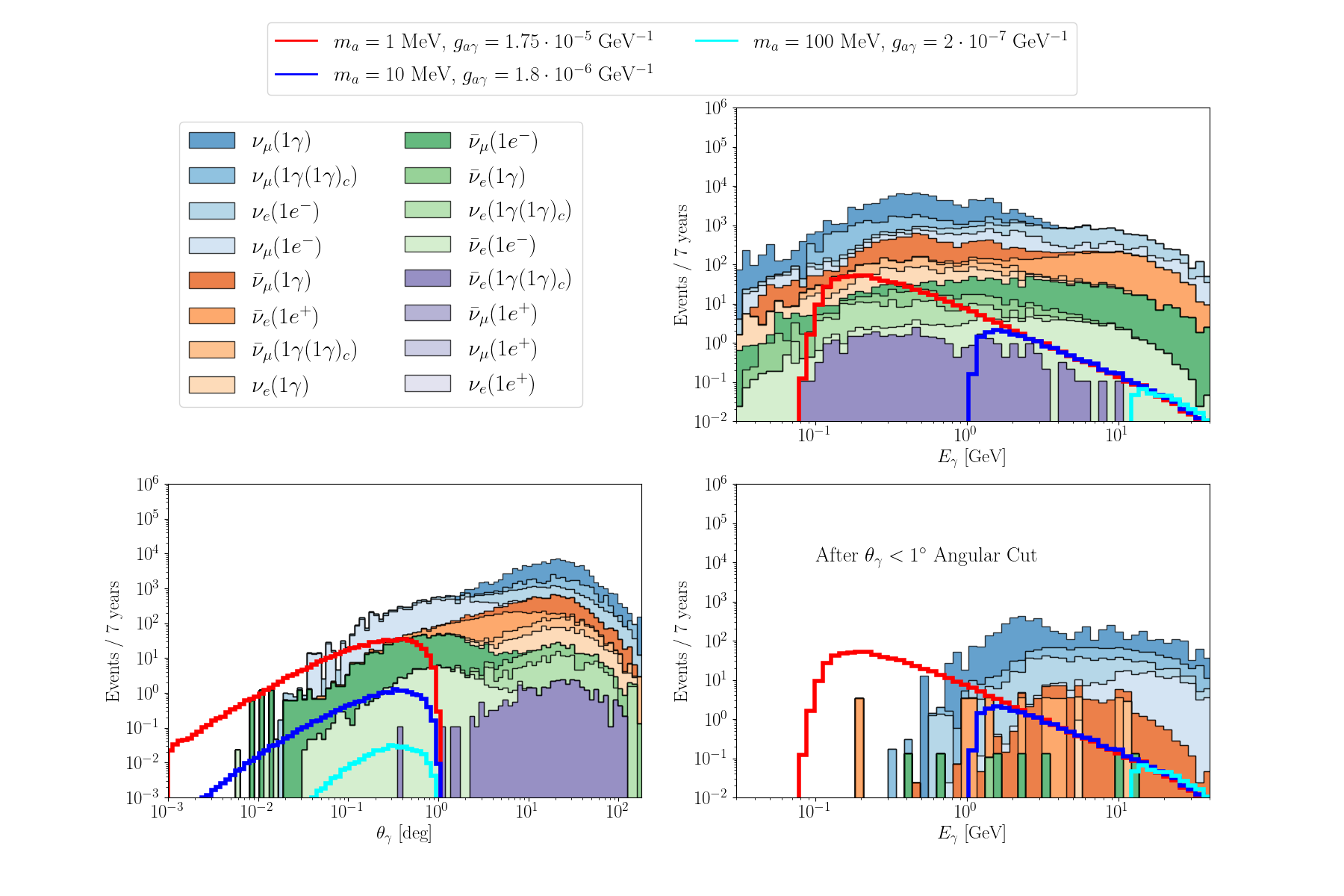}

    \caption{Single photon final state: \textit{Top right}: the ALP energy spectrum from collinear $a \to \gamma \gamma$ decays are shown against the 1$\gamma$ final state backgrounds as well as 1$e^-$ and 1$e^+$ final states with an 18\% mis-ID rate applied. Background labels denoted with $(1\gamma)_c$ stem from photons from true $2\gamma$ events in which one of the photons escaped containment in the fiducial detector volume. \textit{Bottom left}: the ALP signal and background angular spectra for the single shower angle $\theta_\gamma$ taken with respect to the beam axis. \textit{Bottom right}: the same as top right but after the $\theta_\gamma < 1^\circ$ angular cut. Note that the $1e^+$ contributions from the $\bar{\nu}_\mu$, $\nu_\mu$, and $\nu_e$ fluxes (purple shades) are zero.}
    \label{fig:1g0p_backgrounds}
\end{figure}

In Fig.~\ref{fig:1g0p_backgrounds}, we show the signal and background spectra for the $1\gamma$ topology. This signal primarily comes from ALP decays to $\gamma\gamma$ final states that are too collinear to be resolved as separate photons. This occurs most often for lighter ALP masses. For example, at $m_a = 1$ MeV and couplings $g_{a\gamma} \sim \mathcal{O}(10^{-5})$ GeV$^{-1}$ (the red curves), 73\% of the signal flux is forward enough to produce an unresolved single-photon shower; at 10 MeV (dark blue curves), $\sim 10$\%, and at 100 MeV (cyan curves) only $\sim$4\%.
We find that the main kinematic discriminator in this final state is the angle of the single photon-like shower with respect to the beam direction ($\theta_\gamma$); from the bottom left panel of Fig.~\ref{fig:1g0p_backgrounds}, we can motivate a $\theta_\gamma < 1^\circ$ cut. The signal efficiency of this cut is $\geq 99$\%, and does reduce the total background rate from $\mathcal{O}(10^6)$ to $\mathcal{O}(10^3)$. The energy spectrum of the ALP signal after this cut is shown in \cref{fig:1g0p_backgrounds}, bottom right, for the same benchmark ALP mass and coupling combinations.

%%%
\subsection{$e^+ e^-$ Final State}
\label{sec:epem_fs}
%%%

\begin{figure}[t]
    \centering
    \includegraphics[width=1.\textwidth]{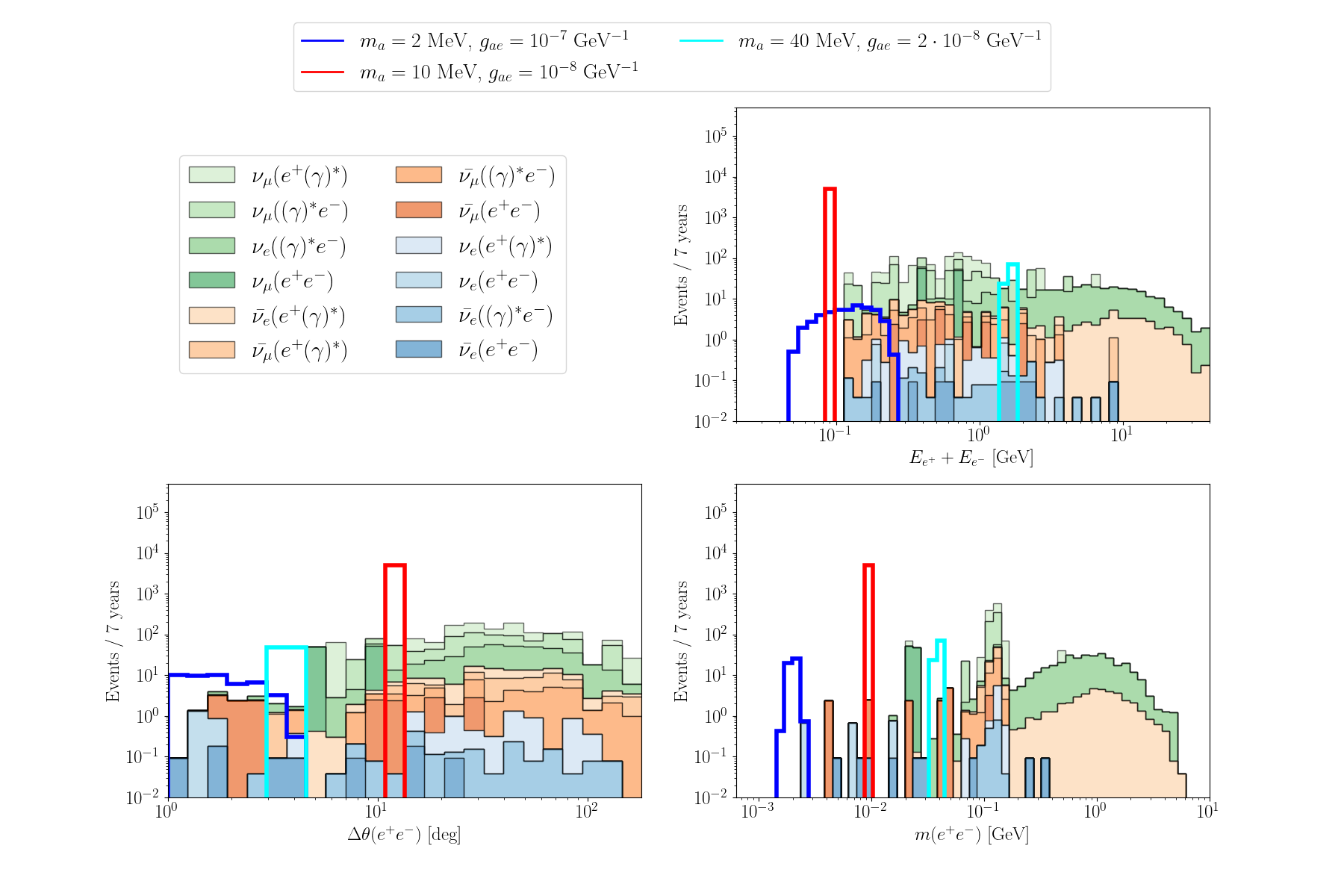}
    
    \caption{Electron-positron final state: energy spectrum (upper right), opening angle (bottom left), and invariant mass (bottom right) distributions are shown for the neutrino-induced background and the ALP signal. Contamination from $1e^+ 1\gamma$ and $1e^- 1\gamma$ backgrounds with one photon mis-identified as an electron or positron are included and indicated with $(\gamma)^*$. The ALP mass and coupling benchmark values are chosen to yield close to a statistically significant number of events at the lower limit of the coupling $g_{ae}$.}
    \label{fig:epem_kinematics}
\end{figure}

ALPs interacting with electrons via the $g_{ae}$ coupling may decay into electron-positron pairs if their masses are $m_a > 2 m_e$, or they may scatter off atomic targets in the detector in the Bethe-Heitler pair production process, $a A \to e^+ e^- A$, to produce a collinear electron-positron pair. Final states with $e^+ e^-$ pairs that can be individually resolved are identified by requiring separation angles of $>1^\circ$ in simulation and individual energies greater than 30 MeV as per Table~\ref{tab:cuts}. Since both particles are resolved, a number of kinematic variables become relevant handles to separate the signal from the background. Like in the case of the $2\gamma$ final state, these include the total energy of the $e^+e^-$ system, the opening angle between the pair, and the invariant mass of $e^+ e^-$ pair, $m(e^+,e^-)$, constructed by summing and squaring their reconstructed four-vectors
\begin{equation}
    m^2(e^+,e^-) = (p_{e^+} + p_{e^-})_\mu(p_{e^+} + p_{e^-})^\mu\,.
\end{equation}
The distributions of reconstructed total energy, opening angle between the electron-positron pair, and reconstructed invariant mass $m(e^+, e^-)$ are shown in Fig.~\ref{fig:epem_kinematics}. Like the 2$\gamma$ final state, the background energy spectrum is falling, while the ALP signal is dominating mostly at low energies. Since the ALP signal from decays is dominated by resonant production ($e^+ e^- \to a$), the energy spectrum is very nearly mono-energetic, peaking at $E_a = m_a^2 / (2 m_e)$. The opening angle of the ALP signal (Fig.~\ref{fig:epem_kinematics}, bottom left) is, of course, larger than $1^\circ$, but does not get larger than $\sim 20^\circ$ for the mass range and couplings (and therefore, boost factors) that we are interested in. These features, like in the 2$\gamma$ case, allow for a good separation between signal and background by placing a cut $\Delta\theta(e^+ e^-) < 20^\circ$ and an invariant mass cut for $|m(e^+,e^-) - m_a| < 0.05 m_a$. 

\begin{figure}[t]
    \centering
    \includegraphics[width=1.0\textwidth]{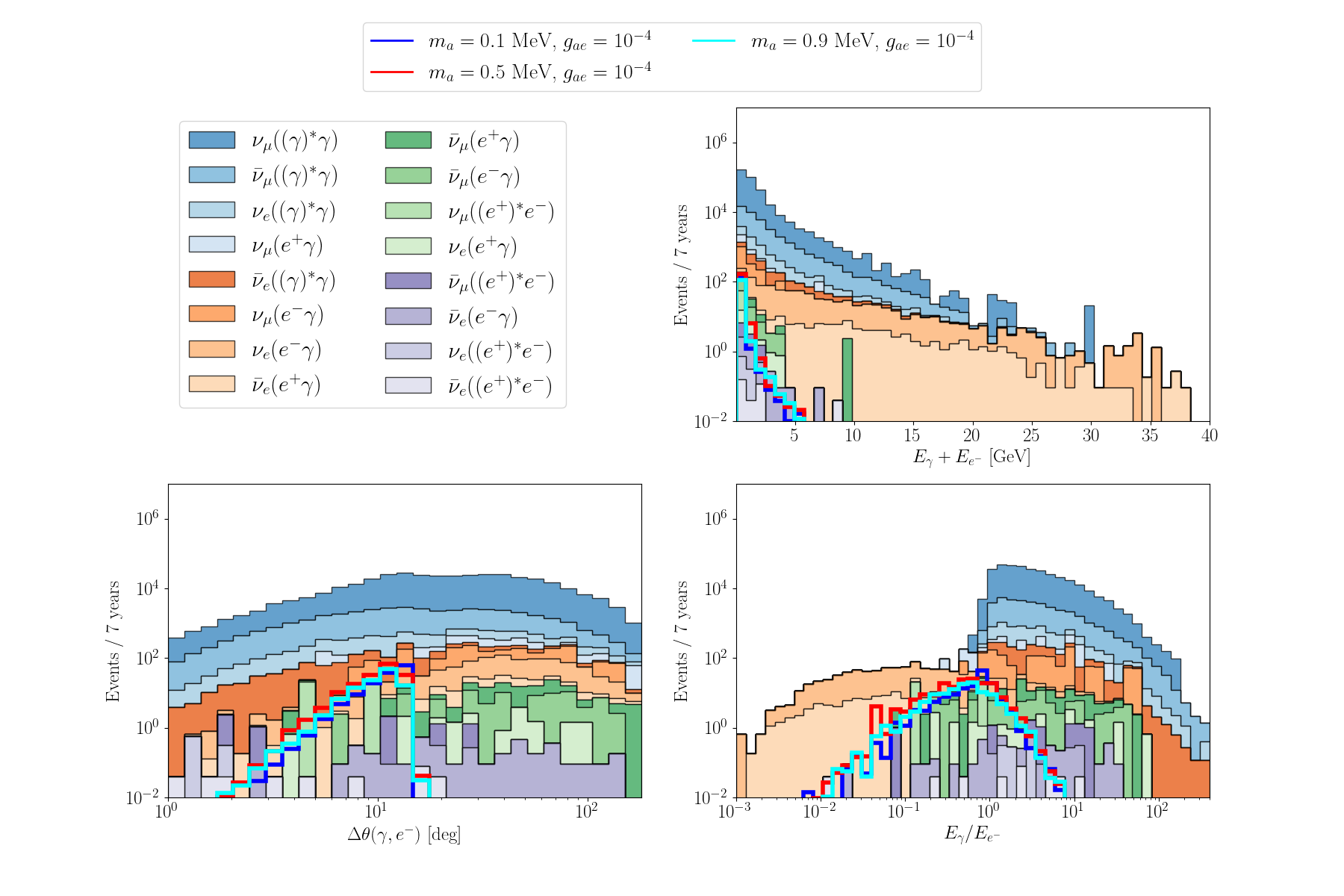}
    \caption{Electron-photon final state: ALP signal distributions plotted against the $1e^- 1\gamma$ neutrino-induced backgrounds for the total combined $E_{e^-} + E_\gamma$ energy (upper right), the opening angle between $e^-$ and $\gamma$ (bottom left), and the energy fraction $E_\gamma/E_{e^-}$ (bottom right). Contamination from $1 e^+ 1\gamma$ is included, as well as $e^+ e^-$ final states with one electron or positron mis-identified as a photon (indicated with $(e^\pm)^*$ in the legend).}
    \label{fig:egamma_signal_vs_bkg}
\end{figure}

\subsection{$e^- \gamma$ Final State}
\label{sec:egamma_fs}
In a similar fashion to the $e^+e^-$ topology, we can construct a set of relevant kinematic variables for the $e^- \gamma$ final state which is used to search for ALPs undergoing inverse Compton scattering via an electron coupling. The electron and photon should be angularly separated to be resolved; we again require $\Delta\theta_{e^- \gamma} > 1^\circ$. To motivate kinematic cuts, we again consider the energies $E_-$ and $E_\gamma$, the angles with respect to the beam axis, $\theta_-$ and $\theta_\gamma$, and their opening angle $\Delta\theta_{e^- \gamma}$. However, instead of reconstructing the invariant mass $m(e^-, \gamma)$, we alternatively consider the energy fraction  $x \equiv E_\gamma / E_{e^-}$. This energy fraction is found to be a better discriminator against the backgrounds given that for pseudoscalar Compton-like scattering, the final state photon tends to carry a small fraction of the energy, while the neutrino-induced backgrounds have a broad energy fraction distribution for the final state photon and electron.

In Fig.~\ref{fig:egamma_signal_vs_bkg}, we compare backgrounds with a typical signal distribution from ALPs produced via the electron coupling $g_{ae}$ and undergoing inverse Compton scattering $a e^- \to \gamma e^-$ in LAr detector. We show this for the total energy (top right), the opening angle between the photon and electron (bottom left), and the energy fraction $x$ (bottom right). As for other final states, an opening angle cut of $\Delta\theta(e^-, \gamma) < 20^\circ$ is clearly motivated. The energy fraction of the final state photon is also well-bounded for the ALP signal, so we will impose a strict cut, keeping only signal and background with $0.1 \leq x \leq 5$. 

\subsection{Summary of Kinematic Cuts}
Finally, we summarize the kinematic cuts we apply to all relevant final states before analyzing the energy spectra.
\begin{itemize}
    \item $\mathbf{\gamma\gamma}$: The relevant cuts are $\Delta\theta(\gamma,\gamma) < 20^\circ$
   and $|m(\gamma,\gamma) - m_a| < 0.05 \times m_a$;
    \item $\mathbf{1\gamma}$: The only relevant kinematic cut is $\theta_\gamma < 1^\circ$;
    \item $\mathbf{{e^+,e^-}}$: We have $\Delta\theta(e^+,e^-) < 20^\circ$ and $|m(e^+,e^-) - m_a| < 0.05 \times m_a$ as the final cuts;
    \item ${\mathbf{e^-\gamma}}$: The cuts are $\Delta\theta(e^-, \gamma) < 20^\circ$ and $0.1 \leq x \leq 5$. 
\end{itemize}

%%%%%%%%%%%%%%%%%%%%%%%%%%%%%%%%%%%%%%%%%%%%%%%%%%%%%%%%%%%%%%%%%%%%%%%%%%%%%%%%%%

%%%%%%%%%%%%%%%%%%%%%%%%%%%%%%%%%%%%%%%%%%%%%%%%%%%%%%%%%%%%%%%%%%%%%%%%%%%%%%%%%%
\section{Sensitivity Projections for DUNE ND-LAr}
\label{sec:sensitivity}

In this section, the forecasted sensitivity for DUNE ND-LAr to ALPs coupling to photons or to electrons via the $g_{a\gamma}$ and $g_{ae}$ couplings, respectively, is presented. These projections are performed assuming a 7 year exposure amounting to $1.029 \times 10^{22}$ POT; 3.5 years in FHC or neutrino mode and 3.5 years in Reverse Horn Current (RHC) or antineutrino mode. For each sensitivity analysis, we construct the following binned Poissonian log-likelihood as our test statistic over $N$ energy bins
\begin{equation}
   \ln L(\vec{\theta}) = \sum_{i=1}^N d_i \ln \big[s_i(\vec{\theta}) + b_i\big] - \big[s_i(\vec{\theta}) + b_i\big] - \ln \big[\Gamma(d_i + 1)\big]\,,    
\end{equation}
for data $d_i$, backgrounds $b_i$, and signal $s_i$ in each energy bin $i$. We take the Gamma function $\Gamma(d_i + 1)$ to analytically continue the usual $d_i !$ for our pseudodata based on the null expectation of neutrino-induced background events, which, after scaling to exposure, are taken as $d_i = b_i \in \mathbb{R}$. To estimate the sensitivity of DUNE ND-LAr, we test the signal hypothesis against the null hypothesis and take the data at its central value from the background-only event rate $d_i = b_i$. The parameter vector $\vec{\theta}$ is taken to be $\vec{\theta}=(m_a, g_{a\gamma})$ or $\vec{\theta}=(m_a, g_{ae})$ depending on which coupling is turned on. This test statistic is also evaluated separately before and after cuts have been made to kinematic variables so that we can compare the sensitivity with and without the cuts. The visible energy is defined as before, taken to be the summed energy of the final state particles after a smearing has been applied (in MC) to model the energy resolution of the detector. 

\begin{figure}[t]
    \centering
    \includegraphics[width=0.497\textwidth]{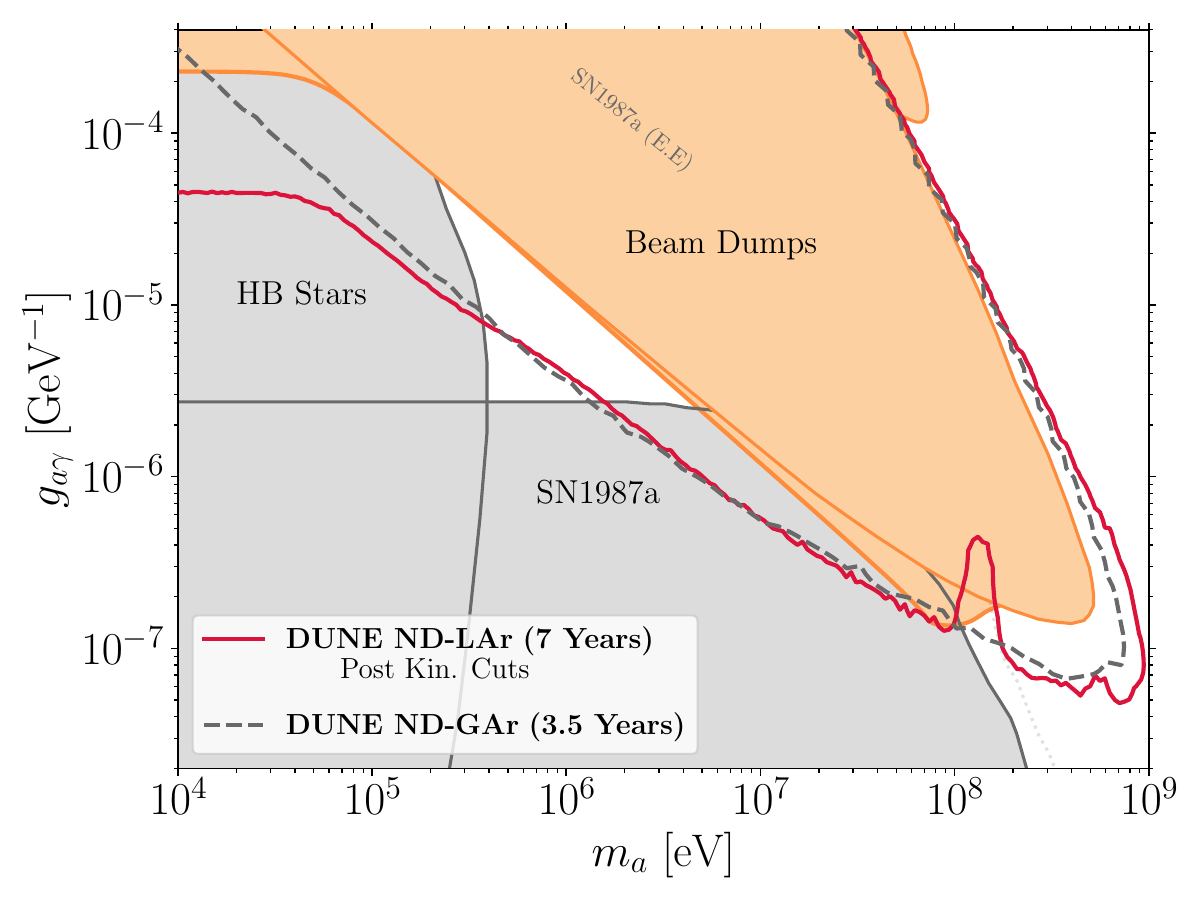}
    \includegraphics[width=0.497\textwidth]{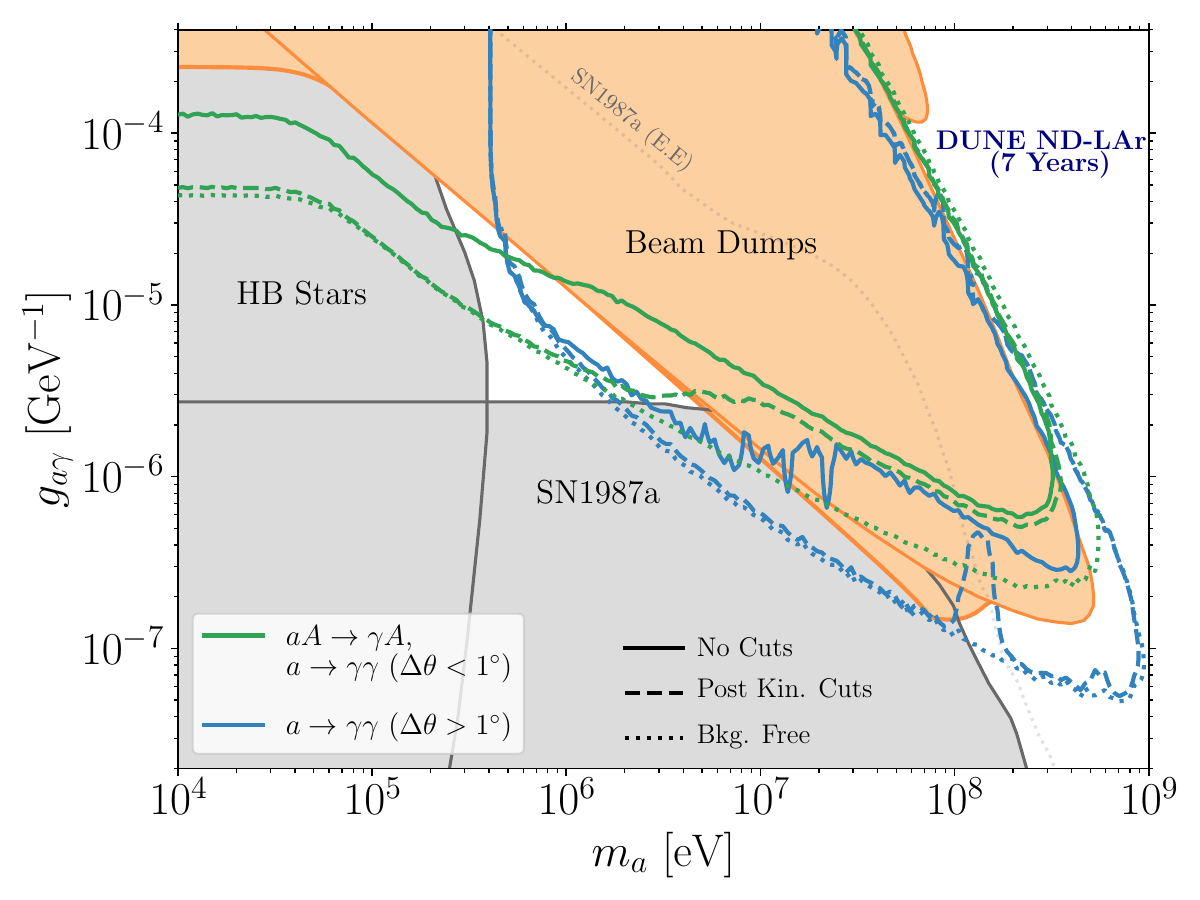}
    \caption{\textit{Left}: Sensitivity reach of the DUNE ND-LAr detector for the ALP-photon coupling $g_{a\gamma}$. Contours of constant $\Delta \ln L$ are drawn at 95\% C.L. for the DUNE ND-LAr sensitivity to combined $1\gamma$ and $2\gamma$ final states with kinematic cuts to reduce backgrounds (red solid). We compare this with the sensitivity to the DUNE ND-GAr with no assumed backgrounds (gray dashed). \textit{Right}: Contours of constant $\Delta \ln L$ for ND-LAr sensitivity are drawn at 95\% C.L. separately for the $1\gamma$ final state (blue) and the $2\gamma$ final state (red) with and without background rejection cuts (dashed and solid, respectively). The $1\gamma$ final state comprises both events from inverse Primakoff scattering $a A \to \gamma A$ and collinear decays $a \to \gamma \gamma$ such that the di-photon pairs are unresolved with $\Delta\theta(\gamma_1, \gamma_2) < 1^\circ$.}
    \label{fig:photon_sens}
\end{figure}

First, we show the sensitivity to ALPs coupled to photons. The sensitivity to ALPs with nonzero $g_{a\gamma}$ coupling and a vanishing $g_{ae}$ coupling is shown in Fig.~\ref{fig:photon_sens}. 
This sensitivity combines the $1\gamma$ and $2\gamma$ final states and their backgrounds listed in Table~\ref{tab:acceptances}, and is constructed by combining the log-Poisson likelihoods as
\begin{equation}
    \ln L_\text{total}(m_a, g_{a\gamma}) = \ln L_{2\gamma} (m_a, g_{a\gamma}) + \ln L_{1\gamma}(m_a, g_{a\gamma})\,.
\end{equation}
We then construct the 95\% confidence level (C.L.) limits by establishing contours of constant delta log-likelihood where
\begin{equation}
    \Delta\ln L(m_a, g_{a\gamma}) \equiv 2 \big[ \text{max}\{ \ln L_\text{total}(m_a, g_{a\gamma}) \} - \ln L_\text{total}(m_a, g_{a\gamma}) \big]\,,
\end{equation}
is equal to 6.18 for a two-parameter scan.

In the left panel of Fig.~\ref{fig:photon_sens}, sensitivity for the  DUNE ND-Lar after the cuts is shown in red. The sensitivity curve follows two primary contributions; one from Primakoff scattering $a + ^{40}{\rm Ar} \to ^{40}{\rm Ar} + \gamma$ which dominates at low masses and larger couplings, where the ALP does not decay before reaching the detector, and another one from decays, which become dominant beyond $m_a \gtrsim 0.1$ MeV, defining the wedge-shaped region in the $m_a - g_{a\gamma}$ plane.

In the right panel of Fig.~\ref{fig:photon_sens}, the effect of the kinematic cuts on the parameter space is visualized. First, focusing on the blue C.L. contours that isolate the resolved $2\gamma$ final states ($\Delta\theta(\gamma,\gamma) > 1^\circ$), we see that the DUNE ND-LAr sensitivity after a 7-year exposure without cuts (solid lines) tests parameter space mostly within regions ruled out by the existing constraints from beam dump experiments (including E137, CHARM, nuCal, and BaBar~\cite{Dolan:2017osp, Aloni:2019ruo}, FASER~\cite{FASER:2024bbl}, and the MiniBooNE beam dump mode~\cite{Capozzi:2023ffu}) except for larger masses $m_a \gtrsim 200$ MeV, where we do find new parameter space sensitivity even without any kinematic cuts. Complementary probes from collider searches test regions of parameter space at higher masses and couplings than the ones shown, e.g. Refs.~\cite{ATLAS:2020hii,CMS:2018erd,Dolan:2017osp,Jaeckel:2015jla,BESIII:2022rzz,BESIII:2024hdv}. The sensitivity contour without any background cuts exhibits some noisy behavior in the $m_a \sim 10$ MeV regime due to low-statistics regions in the background distribution. The effect of the angular cuts and invariant mass cut improves the sensitivity (dashed lines) by a factor of up to $100\%$ in the $m_a = 20-100$ MeV range, where the ALP energy spectrum shifts into a background-dominated region (see Fig.~\ref{fig:2g_kinematics}) and the invariant mass cut plays a bigger role in background removal. The effect of the cuts allows for more sensitivity at lower couplings, beyond the existing laboratory constraints from the beam dump experiments. However, the invariant mass cut loses background rejection power when $m_a \simeq m_\pi, m_\eta$, where neutrino-induced coherent pion and eta meson production dominates the invariant mass spectrum, as seen in Fig.~\ref{fig:2g_kinematics}. The astrophysical constraints are comprised of limits from supernovae~\cite{Lucente:2020whw} and HB star cooling~\cite{Carenza:2020zil}, as well as the measurement of the explosion energy of SN1987A (gray dotted line, shown for $10^51$ erg energy deposition), which lies in tension with the so-called ``cosmological triangle region'' unless the star cooling process is significantly different from the standard picture~\cite{Caputo:2021rux,Fiorillo:2025yzf}, see also refs.~\cite{Khoury:2003aq,Masso:2005ym,Masso:2006gc,Dupays:2006dp, DeRocco:2020xdt, Mohapatra:2006pv,Brax:2007ak,Jaeckel:2006xm}. For unresolved ALP decays ($\Delta \theta(\gamma,\gamma) < 1^\circ$) and inverse Primakoff scattering (green C.L. contours), both of which are reconstructed as single showers, the angular cut of the single photon with respect to the beam direction is highly efficient such that the projected sensitivity after kinematic cuts is nearly background-free.

When comparing the background-free sensitivity in LAr to the background-free sensitivity calculated for the gaseous argon (GAr) detector running in target-mode (see ref.~\cite{Brdar:2020dpr}), we find that the sensitivity in LAr is worse by a factor of 5 in the coupling $g_{a\gamma}$. This was found to be due in part to an overestimation of the target photon flux from a normalization error in ref.~\cite{Brdar:2020dpr}. Nevertheless, the sensitivity toward the upper-right region (called ``beam-dump ceiling''~\cite{Dutta:2023abe}), where ALPs decay rather promptly, is insignificantly affected by this overestimation due to the fact that the coupling dependence here is effectively exponential~\cite{Kim:2024vxg}. The GAr setup assumed in ref.~\cite{Brdar:2020dpr} was also larger in effective volume and would enjoy a suppressed neutrino-induced background rate due to the lower density, see also ref.~\cite{Coloma:2023oxx}. The relatively weaker sensitivity found in the current study, with a more sophisticated treatment of the signal flux and background rates, is mainly impacted in the ``cosmological triangle'' region, but this region is now known to be likely ruled out by supernova limits~\cite{Caputo:2021rux} (and potentially and partially reactors, as claimed recently~\cite{Park:2024omu}). In the high-mass limit, however, we still find a good advance in sensitivity to ALPs beyond the existing beam dump limits up to nearly 1 GeV in ALP mass. Since the ND-LAr detector would likely be present in both phases of DUNE~\cite{DUNE:2024wvj}, while ND-GAr may only be present in the second phase, our emphasis remains on the LAr configuration to achieve the best sensitivity to new physics parameter space in the near term.

\bigskip

\begin{figure}[t]
    \centering
    \includegraphics[width=0.497\textwidth]{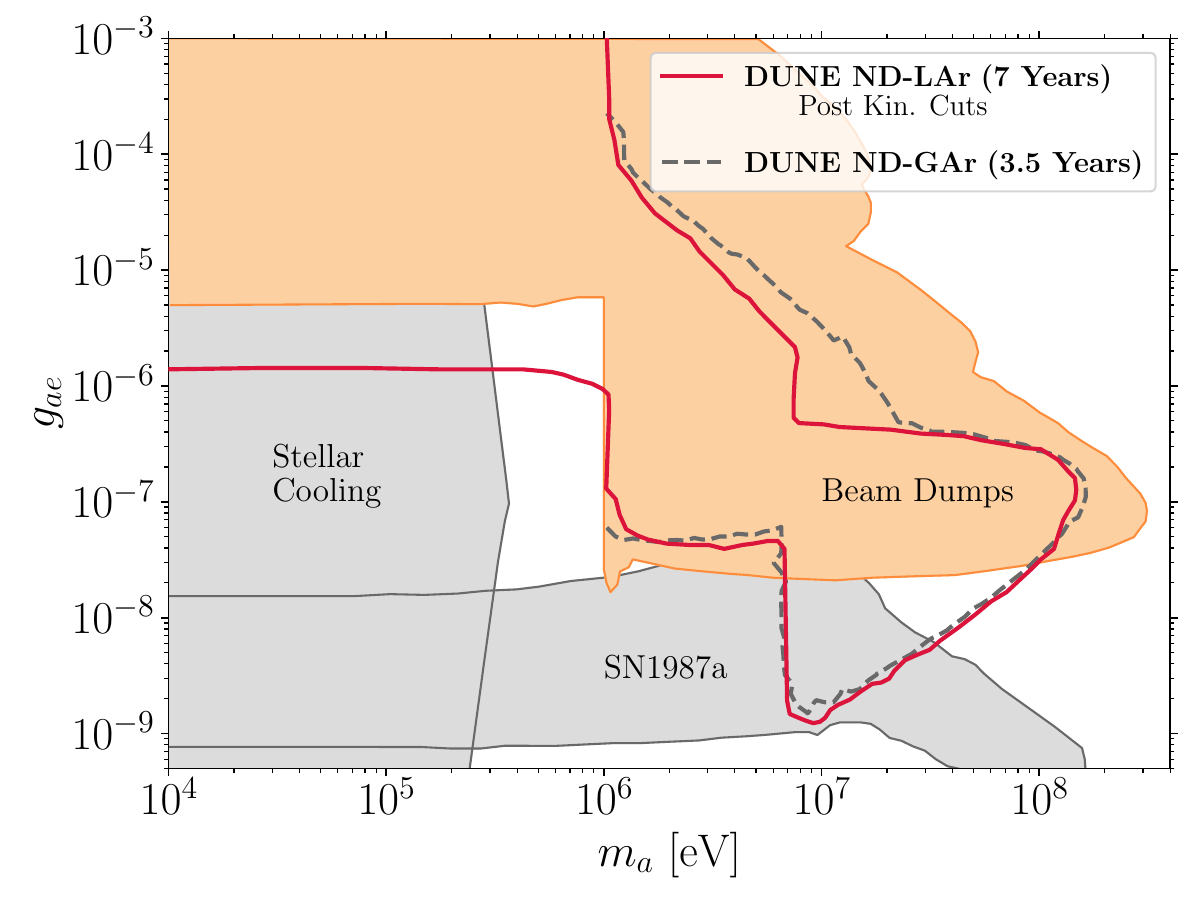}
    \includegraphics[width=0.497\textwidth]{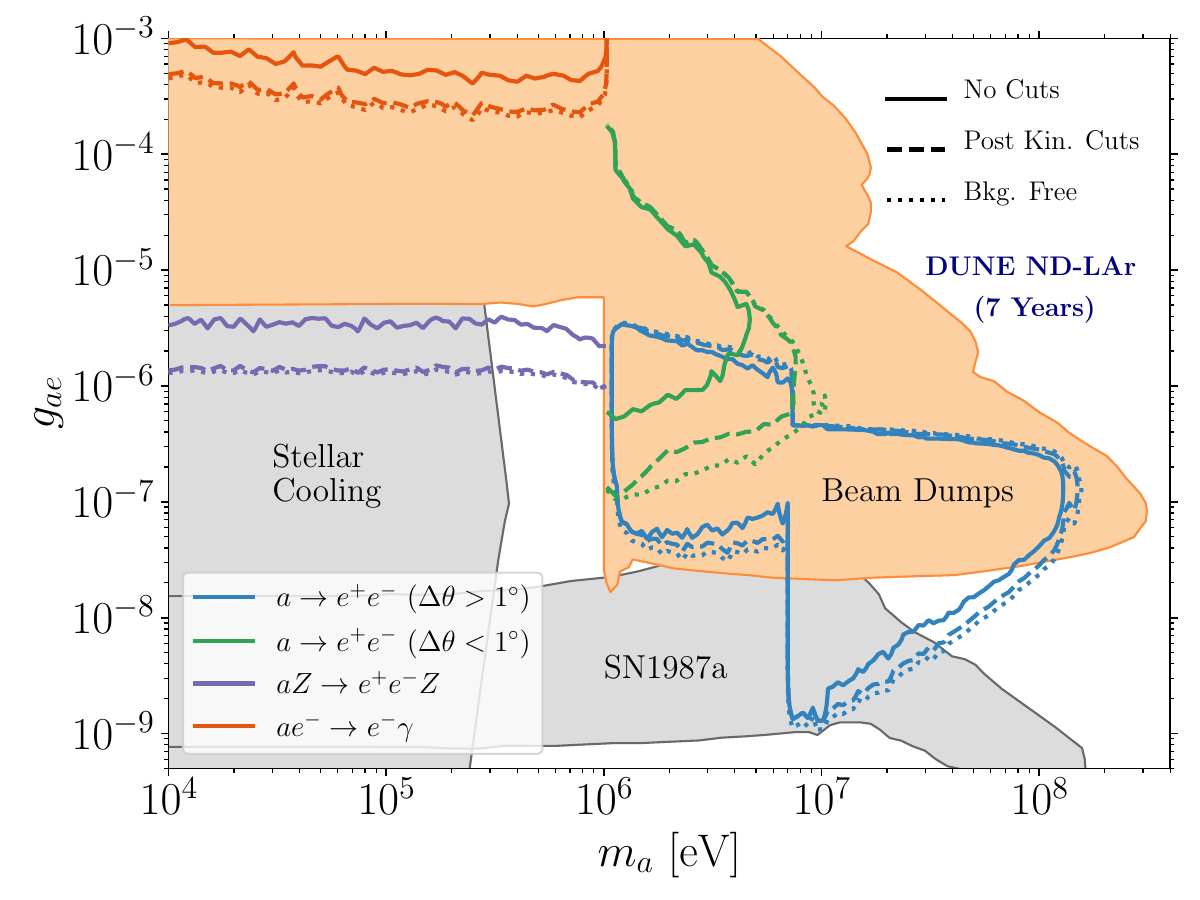}
    
    \caption{\textit{Left}: Sensitivity reach of the DUNE ND-LAr detector after kinematic cuts (red solid) and the background-free ND-GAr detector (gray dashed) to the ALP-electron coupling $g_{ae}$. Contours of constant $\Delta \ln L$ are drawn at 95\% C.L. \textit{Right}: Sensitivity reach of the DUNE ND-LAr detector to the ALP-electron coupling $g_{ae}$ separated channel-by-channel. Contours of constant $\Delta \ln L$ are drawn at 95\% C.L., before and after kinematic cuts, separately for the $1\gamma1e^-$ final state, the $1e^+1e^-$ final state, and the collinear $e^+e^-$ or $1(\gamma)^*$ final state. }
    \label{fig:electron_sens_combined}
\end{figure}

Next, we present the sensitivity of the DUNE ND-LAr to ALPs with electron couplings, keeping $g_{a\gamma} = 0$. For this sensitivity, we combine three topologies of the final state. We consider  $e^+ e^-$ final state, which includes ALP decays, $a \to e^+ e^-$, and pair production scattering $a A \to e^+ e^- A$. We also consider $1(\gamma)^*$ final state, effectively created from collinear $e^+ e^-$ pairs that are mis-identitified as a single $\gamma$ when their opening angle is less than 1$^\circ$. Finally, we consider the $1\gamma 1e^-$ final state due to Compton-like scattering $a e^- \to \gamma e^-$ off atomic electrons in the liquid argon. 

The sensitivity curves for the $g_{ae}$ coupling are shown in the left panel of Fig.~\ref{fig:electron_sens_combined}, while the right panel shows the sensitivity channel-by-channel. In the sub-MeV mass range, where ALP decays to $e^+ e^-$ are kinematically forbidden, we find that the most sensitive channels for ALP detection stem from electron-positron pair conversion $a A \to e^+ e^- A$, shown in purple (unless one considers electron loop-induced $\gamma\gamma$ decays, in which case beam dump constraints would be modified~\cite{Bauer:2020jbp}, and the $\gamma\gamma$ channel would shift the sensitivity contour down in couplings between $2 m_e$ and $m_a \simeq 0.1$ MeV where the lifetime is long enough such that scattering into $e^+ e^-$ pair conversion dominates again~\cite{Liu:2017htz,CCM:2021jmk}). In the $m_a > 2 m_e$ mass range, decays to resolved $e^+ e^-$ pairs (blue) dominate the sensitivity. In the case of pair conversion, the final state $e^+ e^-$ pair is highly collinear and merged into a single shower, so the most effective kinematic cut to reduce background is on the angle of the $e^+ e^-$ pair with respect to the beam axis (discussed in Section~\ref{sec:single_gamma_fs}). This accounts for a nearly five-fold increase in the sensitivity to the coupling $g_{ae}$ and results in an almost background-free analysis after the kinematic cuts. For the resolved case, the sensitivity to $e^+ e^-$ from ALP decays is best enhanced from the combination of opening angle and invariant mass cuts discussed in Section~\ref{sec:epem_fs}. Note from the right panel of Fig.~\ref{fig:electron_sens_combined} the sudden increase in sensitivity at $m_a \simeq 7$ MeV due to the resonant production channel energy, $E_a \sim m_a^2 / (2 m_e)$, which can only give rise to electron-positron pairs above the detector energy thresholds of $30$ MeV, or when $m_a^2 \gtrsim 2 \times 30$ MeV$^2$.

With these cuts, DUNE ND-LAr sensitivity reaches $g_{ae} \simeq 10^{-9}$ level, testing parameter space beyond the existing beam dump bounds (combined from E137~\cite{PhysRevD.38.3375,Andreas:2010ms}, Orsay~\cite{Bechis:1979kp}, E141~\cite{Riordan:1987aw}, E774~\cite{Bross:1989mp}, NA64~\cite{NA64:2021ked,Andreev:2021fzd,Gninenko:2017yus}, and the CCM120 engineering run~\cite{CCM:2021jmk}) and entering regions so far excluded only by astrophysics (limits from SN1987A~\cite{Lucente:2021hbp}, which may also be further constrained from explosion energy deposition~\cite{Fiorillo:2025sln}, and stellar cooling~\cite{Hardy:2016kme}). Complementary probes from collider searches at higher masses and couplings than the ones shown here can be found in Refs.~\cite{Bauer:2017ris, Eberhart:2025lyu, Alimena:2025kjv}.

%%%%%%%%%%%%%%%%%%%%%%%%%%%%%%%%%%%%%%%%%%%%%%%%%%%%%%%%%%%%%%%%%%%%%%%%%%%%%%%

%%%%%%%%%%%%%%%%%%%%%%%%%%%%%%%%%%%%%%%%%%%%%%%%%%%%%%%%%%%%%%%%%%%%%%%%%%%%%%%
\section{Conclusions}
\label{sec:conclusion}
In this work, we have investigated the neutrino-induced backgrounds relevant to new physics signals with $e^+ e^-$, $2\gamma$, $1 e^\pm 1 \gamma$, and $1\gamma$ final states. We have modeled these backgrounds by simulating the interaction of neutrinos from the DUNE beam with liquid argon using \texttt{\texttt{GENIE}}, accounting for reconstruction effects on the event selection, particle mis-identification, and detector smearing effects. Long-lived ALPs that couple to electrons or photons give rise to the above final states, and by examining the kinematics of the final state, we motivated a set of kinematic cuts to reduce the background rates.

We expect that our comprehensive background analysis for various signal topologies from ALPs at the DUNE near detector can serve as a litmus test for the DUNE BSM program. We have shown that the electromagnetic signatures considered in this work can be separated from their neutrino-induced backgrounds, using information on the opening angle, angle with respect to the beam axis, energy ratios, and invariant mass. We expect DUNE to test new parameter space beyond the existing laboratory bounds with a 7-year exposure. Namely, for ALPs coupled to photons, from unresolved $a\to \gamma \gamma$ decays and inverse Primakoff scattering producing single showers in the detector, the entirety of the $m_a \sim 1$ MeV parameter space, the so-called ``cosmological triangle'', can be probed. For larger masses $m_a = 0.1-1$ GeV, resolved $a\to \gamma \gamma$ decays can probe couplings below the $g_{a\gamma} = 10^{-7}$ GeV$^{-1}$ level, beyond the existing beam dump constraints.  For ALPs coupled to electrons, we forecast sensitivity in the $m_a = 20-100$ MeV mass range from $a \to e^+ e^-$ decays, expanding upon the existing beam dump constraints and testing the SN1987A bound. In the $m_a < 2 m_e$ mass range, an improvement of half an order-of-magnitude in the electron coupling $g_{ae}$ can be achieved. In both of these cases, while the $e^+ e^-$ background is already quite manageable, almost 100\% background reduction can be achieved with angular kinematic cuts.

Since many other scenarios in which BSM physics can show up in beam target facilities also have electromagnetic signatures, e.g., dark photons~\cite{Berryman:2019dme}, extended Higgs sector scalars~\cite{Berryman:2019dme}, heavy neutral leptons~\cite{Ballett:2019bgd,Capozzi:2024pmh}, dark matter scattering involving mutilepton final states~\cite{Kim:2019had,Dutta:2024nhg} or multi-particle dark matter with upscattering~\cite{Kim:2016zjx,Chatterjee:2018mej,Kim:2020ipj,DeRoeck:2020ntj,Batell:2021ooj}, our results for the ALP case with electromagnetic couplings should translate easily to other scenarios and provide a good sense of how background-impacted they may be.
%%%%%%%%%%%%%%%%%%%%%%%%%%%%%%%%%%%%%%%%%%%%%%%%%%%%%%%%%%%%%%%%%%%%%%%%%%%%%%%

%%%%%%%%%%%%%%%%%%%%%%%%%%%%%%%%%%%%%%%%%%%%%%%%%%%%%%%%%%%%%%%%%%%%%%%%%%%%%%%
\section*{Acknowledgments}
We are grateful to Pablo Alzas, Daniel Cherdack, Josu Hern\'{a}ndez-Garc\'{i}a, Justo Mart\'{i}n-Albo, and Vishvas Pandey for their helpful discussions on resolution effects and particle identification capabilities of LArTPC technology and to Chris Marshall for the helpful feedback.
The work of VB is supported by the U.S. Department of Energy (DOE) Grant No. DE-SC0025477. The work of DK is supported in part by the DOE Grant No. DE-SC0010813. The work of AT is supported in part by the U.S. DOE grant DE-SC0010143. IMS is supported by the U.S. DOE Office of Science under award number DE-SC0020262. The work of ZT is supported by Pitt PACC and CERN's Theoretical Physics department. The work of JY and WYJ is supported in part by the U.S. DOE under Grant No. DE-SC0011686. The work of BD is supported by the U.S. DOE Grant~DE-SC0010813. We would like to thank the Center for Theoretical Underground Physics and Related Areas (CETUP*) and the Institute for Underground Science at Sanford Underground Research Facility (SURF) for providing a conducive environment during the
2025 summer workshop. We also acknowledge discussions at the NEAT workshop at Colorado State University in May 2025 which augmented this work. The code used for this research is made publicly available through~\gitlink under CC-BY-NC-SA~\cite{alplib}.
%%%%%%%%%%%%%%%%%%%%%%%%%%%%%%%%%%%%%%%%%%%%%%%%%%%%%%%%%%%%%%%%%%%%%%%%%%%%%%%

%%%%%%%%%%%%%%%%% APPENDIX %%%%%%%%%%%%%%%%%%%%%%
\appendix

\section{Pair Production Cross Section Integration}
\label{app:pair_prod}
To compute the rate of $e^+ e^-$ pair production from ALP scattering, we calculate the matrix element for the process $a A \to e^+ e^- A$, where $A$ is the combined particle wavefunction of the electron cloud and nuclear electric charge of an atom ($^{40}$Ar in this case). See also ref.~\cite{Arias-Aragon:2024gdz} for another presentation of this cross section and its relevance for solar and supernovae axion searches.

\begin{figure}[h]
 \centering
        \begin{tikzpicture}
              \begin{feynman}
         \vertex (o1);
         \vertex [left=1.4cm of o1] (i1) {\(a(k)\)};
         \vertex [right=1.4cm of o1] (f1){\(e^-(p_-)\)};
         \vertex [below=1.4cm of o1] (o2);
         \vertex [right=1.4cm of o2] (f2){\(e^+(p_+)\)};
         \vertex [below=1.4cm of o2] (o3);
         \vertex [left=1.4cm of o3] (i2) {\(N(\ell)\)};
         \vertex [right=1.4cm of o3] (f3) {\(N(\ell^\prime)\)};

         \diagram* {
           (i1) -- [scalar] (o1),
           (f2) -- [fermion1] (o2) -- [fermion1] (o1) -- [fermion1] (f1),
           (o2) -- [boson, edge label={\(\gamma (q)\)}] (o3),
           (i2) -- [fermion1] (o3),
           (o3) -- [fermion1] (f3),
         };
        \end{feynman}
       \end{tikzpicture}
       \begin{tikzpicture}
              \begin{feynman}
         \vertex (o1);
         \vertex [left=1.4cm of o1] (i1) {\(a(k)\)};
         \vertex [below right=1.4cm of o1] (f1){\(e^+(p_+)\)};
         \vertex [below=1.4cm of o1] (o2);
         \vertex [above right=1.4cm of o2] (f2){\(e^-(p_-)\)};
         \vertex [below=1.4cm of o2] (o3);
         \vertex [left=1.4cm of o3] (i2) {\(N(\ell)\)};
         \vertex [right=1.4cm of o3] (f3) {\(N(\ell^\prime)\)};

         \diagram* {
           (i1) -- [scalar] (o1),
           (f1) -- [fermion1] (o1) -- [fermion1] (o2) -- [fermion1] (f2),
           (o2) -- [boson, edge label={\(\gamma (q)\)}] (o3),
           (i2) -- [fermion1] (o3),
           (o3) -- [fermion1] (f3),
         };
        \end{feynman}
       \end{tikzpicture}
    \caption{Tree-level diagrams for axion pair production.}
    \label{fig:pair_prod_feynman}
\end{figure}
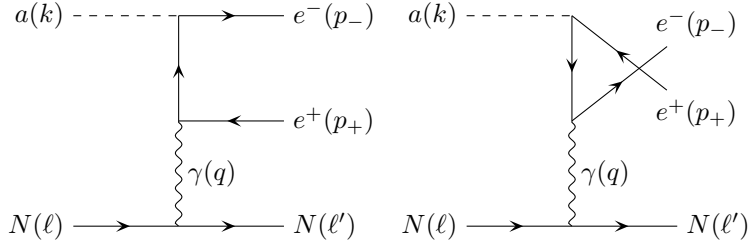

We calculate the matrix element for this process scattering on a single fermion at a time, with momenta assigned as $a(k) + N(\ell) \to e^+(p_+) + e^-(p_-) + N(\ell^\prime)$, analogous to the pair conversion process in which the SM photon converts to an electron-positron pair in the presence of the strong electromagnetic field of a nucleus. This is the dominant mechanism for photon absorption for $E_\gamma \gtrsim 10$ MeV, and one can expect a similarly large cross-section for ALP scattering at MeV energies. The 4-momenta in the laboratory frame are defined as follows
\begin{align}
    k^\mu :& \, \, \,  (E_a, 0, 0, p_a) \nonumber \\
    p_+^\mu :& \, \, \,  (E_+, p_+\cos\phi_+\sin\theta_+, p_+\sin\phi_+\sin\theta_+, p_+\cos\theta_+) \nonumber \\
    p_-^\mu :& \, \, \,  (E_-, p_-\cos\phi_-\sin\theta_-, p_-\sin\phi_-\sin\theta_-, p_-\cos\theta_-) \nonumber \\
    \ell^\mu :& \, \, \,  (m_N, 0, 0, 0) \nonumber \\
    \ell^{\prime\mu} :& \, \, \,  (m_N, \vec{q}) \nonumber \,.
\end{align}
This process will be considered at both non-relativistic and fully relativistic regimes; however, we have made the approximation $\ell^{\prime 0} = m_N$, i.e., that the energy transfer to the nucleus is negligible since $T = q^2 / (2 m_N)$ is expected to be small with $m_N \gtrsim 10^4$ MeV. The matrix element can then be written using the Feynman rules
\begin{align}
    \mathcal{M}_1 &= \Bar{u}(p_-) (ie\gamma^\mu)\bigg( \dfrac{i(\slashed{k} - \slashed{p}_+ + m)}{(k - p_+)^2 - m_e^2} \bigg)(i g_{ae} \gamma^5)  v(p_+) \bigg(\dfrac{-i}{q^2} \bigg) \Bar{u}(\ell^\prime)(ie\gamma_\mu) u(\ell) \nonumber \\
    &= \dfrac{-i g_{ae} e^2}{q^2 ((k-p_+)^2 - m_e^2)} [\Bar{u}(p_-)\gamma^\mu (\slashed{k} - \slashed{p}_+ + m)\gamma^5 v(p_+)][\Bar{u}(\ell^\prime)\gamma_\mu u(\ell)] \,,
\end{align}
\begin{align}
    \mathcal{M}_2 &= \Bar{u}(p_-) (i g_{ae} \gamma^5)\bigg( \dfrac{i(\slashed{k} - \slashed{p}_- + m)}{(k - p_-)^2 - m_e^2} \bigg)(ie\gamma^\mu)  v(p_+) \bigg(\dfrac{-i}{q^2} \bigg) \Bar{u}(\ell^\prime)(ie\gamma_\mu) u(\ell) \nonumber \\
    &= \dfrac{-i g_{ae} e^2}{q^2 ((k-p_-)^2 - m_e^2)} [\Bar{u}(p_-)\gamma^5 (\slashed{k} - \slashed{p}_- + m)\gamma^\mu v(p_+)][\Bar{u}(\ell^\prime)\gamma_\mu u(\ell)] \,.
\end{align}
The total squared amplitude is given by $|\mathcal{M}|^2 = \mathcal{M}_1^\dagger \mathcal{M}_1 + \mathcal{M}_2^\dagger \mathcal{M}_2 + 2 \mathcal{M}_1^\dagger \mathcal{M}_2$. Squaring and summing (averaging) over final (initial) state spins then gives us the total matrix element.
Again, the expression for the differential cross-section is inspired by a similar setup for computing SM photon conversion to an electron-positron pair~\cite{PhysRev.93.768,bethe1953}
\begin{equation}
\label{}
    d\sigma^{e^+e^-} = \dfrac{1}{4 E_a m_N v_a}  \braket{|\mathcal{M}|^2} \dfrac{d^3 p_+}{(2\pi)^3 2E_+} \dfrac{d^3 p_-}{(2\pi)^3 2E_-} \dfrac{d^3 q}{(2\pi)^3 2m_N} (2\pi)^4 \delta^4 (\ell^\prime + p_+ + p_- - k_a - \ell) \,.
\end{equation}

\begin{figure}
    \centering
    \includegraphics[width=0.9\textwidth]{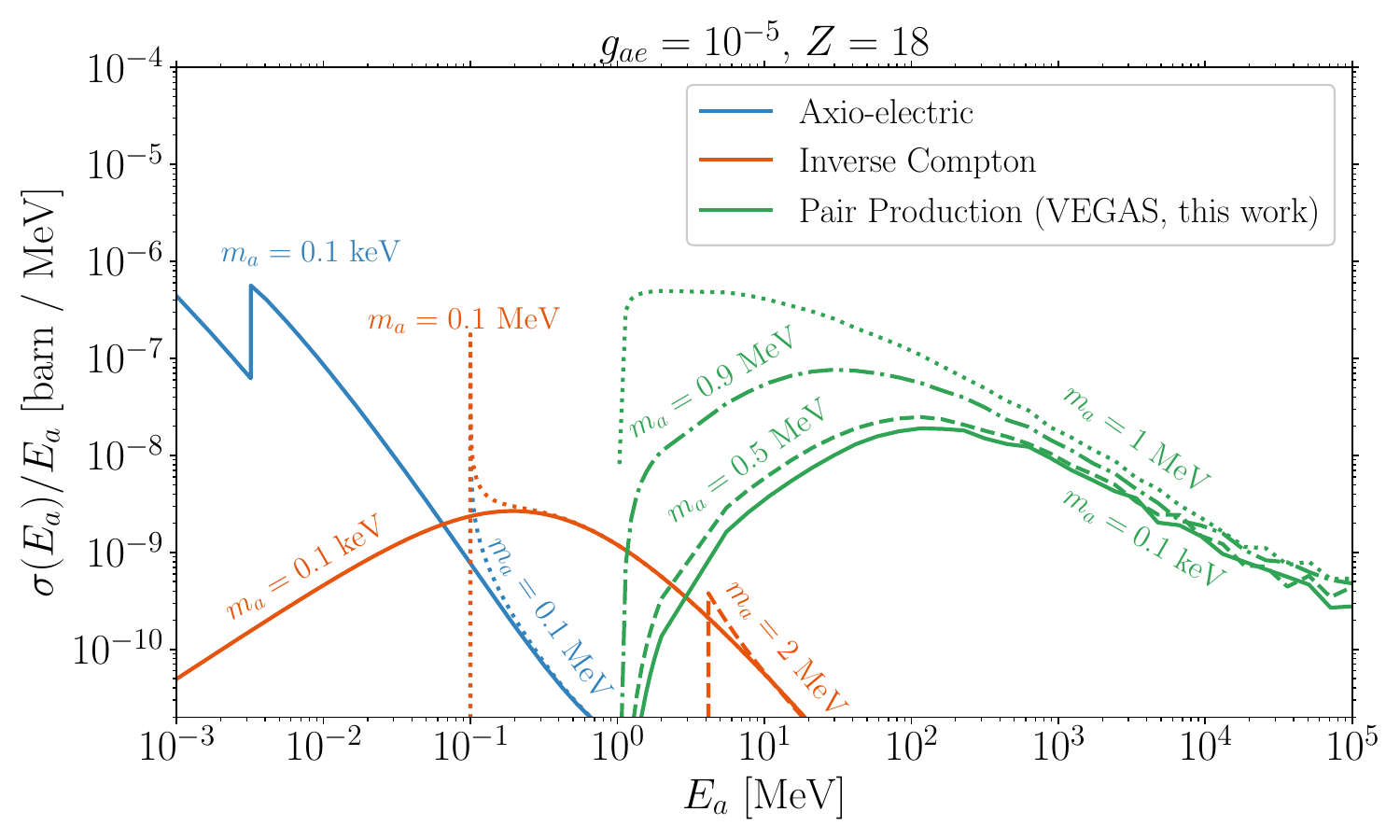}
    \caption{Pair production $(a \, A \to e^+ \, e^- \, A)$ scattering cross-section, integrated with VEGAS, compared to axio-electric and inverse Compton scattering channel cross-sections (in the figure, we show cross sections divided by the ALP energy $E_a$).}
    \label{fig:det_xs_comparisons}
\end{figure}
The total cross-section is shown in Fig.~\ref{fig:det_xs_comparisons} for several benchmark ALP masses (green lines). We compare this cross-section with the rates of other ALP scattering channels; the cross-section for axio-electric ionization and excitation~\cite{Derevianko:2010kz} is shown in blue, while the cross-section for inverse Compton scattering is shown in red. Because of the asymptotic $1/s$ behavior of inverse Compton scattering, pair production is left as the dominant scattering channel for high-energy ALPs like those that could be produced by the DUNE beam. 

\section{Form Factors}
In coherent scattering processes which involve the photon in the $t$-channel mediator to an atomic scattering target, like those in ALP Primakoff scattering or pair production, the low momentum transfer ($q$) limit is faced with a screening of the atomic nucleus which is realized as a suppression in the form factor. One can use the atomic form factor with a point-like nucleus~\cite{RevModPhys.46.815}
\begin{equation}
    G_{\rm el}^2(t) = Z^2 \bigg( \frac{a^2 t}{1 + a^2 t} \bigg)^2 \,,
\end{equation}
for $a \equiv 111 Z^{-1/3} / m_e$ and we take $t \equiv q^2$. However, in the limit of large momentum transfer, the resolution of the nucleus will eventually suppress the coherence. One can use the form factor, which includes the nuclear dipole, e.g., from refs.~\cite{Arias-Aragon:2024gdz, PhysRevD.8.3109, RevModPhys.46.815}. We take $G_{\rm el}^2(t) \to  G_{\rm el}^2(t) \times 1/(1 + t/d)^2$
where $d = 0.164 A^{-2/3}$ GeV$^2$, $m_e$ is the electron mass, and $A = N + Z$ is the number of neutrons plus protons. 

\begin{figure}[t]
    \centering
    \includegraphics[width=0.7\linewidth]{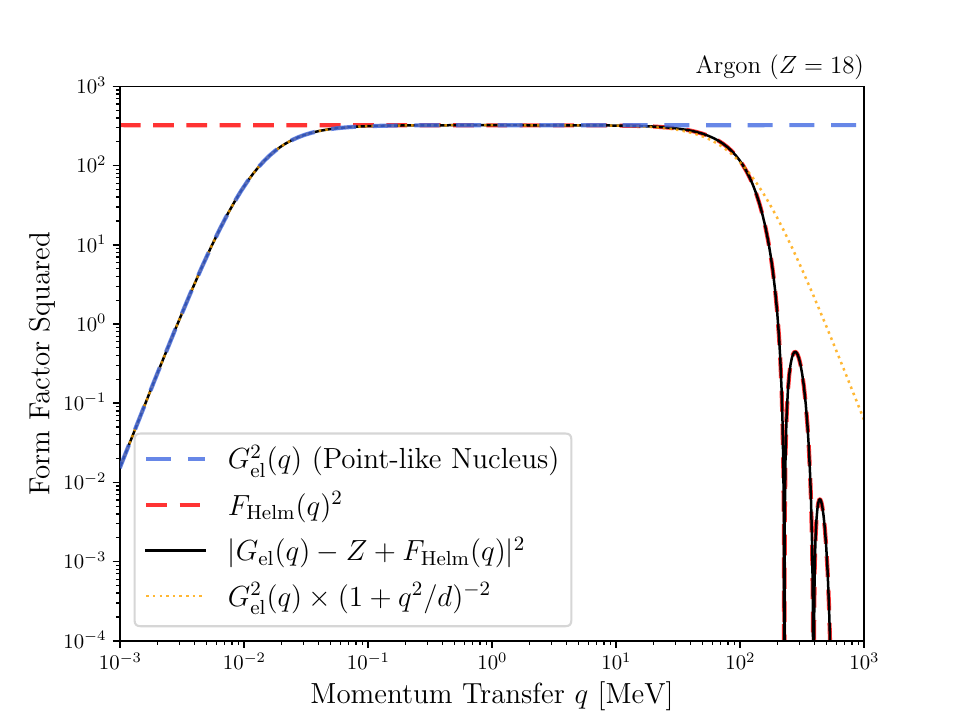}
    \caption{A comparison between form factors for an argon ($Z=18$) nucleus. The Helm parameterization of the nuclear form factor is shown (red, dashed) as well as the atomic form factor with a point-like nucleus (blue, dashed). The elastic form factor with nuclear dipole (orange, dashed) can then be compared with the Helm + elastic form factor (black, solid), constructed by subtracting off the point-like nucleus contribution of charge $Z$ to the form factor before squaring.}
    \label{fig:form_factors}
\end{figure}

Alternatively, one could approximately capture the resolution of the nucleus by combining the atomic form factor above with the Helm parametrization of the nuclear charge density (see e.g., refs.~\cite{Engel:1992bf, Chen:2011xp}) in an \textit{ad hoc} way by subtracting off the point-like contribution and adding back in the Helm form factor
\begin{equation}
\label{eq:ff_with_helm}
    F^2_A(t) = |G_{\rm el}(q) - Z + F_{\rm Helm}|^2 \,,
\end{equation}
where $F_{\rm Helm} = 3 Z j_1(q, r_N) \exp(-q^2 s^2 / 2) / (q \,  r_N)$, $j_1$ is the Bessel function of the first kind, and $r_N$ is the neutron skin radius.\footnote{We utilize the Helm parameterization as an approximation of the nuclear charge form factor. We take $r_N = \sqrt{(1.23 A^{1/3} - 0.6)^2 - 5 s^2 + (7/3) \pi^2 b^2}$ fm with $s=0.9$ fm and $b=0.52$ fm.}
The difference between eq.~\eqref{eq:ff_with_helm} and the nuclear dipole at large momentum transfer can be seen in Fig.~\ref{fig:form_factors}. In the present work, we conservatively prefer to use eq.~\eqref{eq:ff_with_helm} to describe the form factor for Primakoff ($\gamma A \leftrightarrow a A$) and pair production ($a A \to e^+ e^- A$).

\section{Timing Spectra}

\begin{figure}
    \centering
    \includegraphics[width=0.75\textwidth]{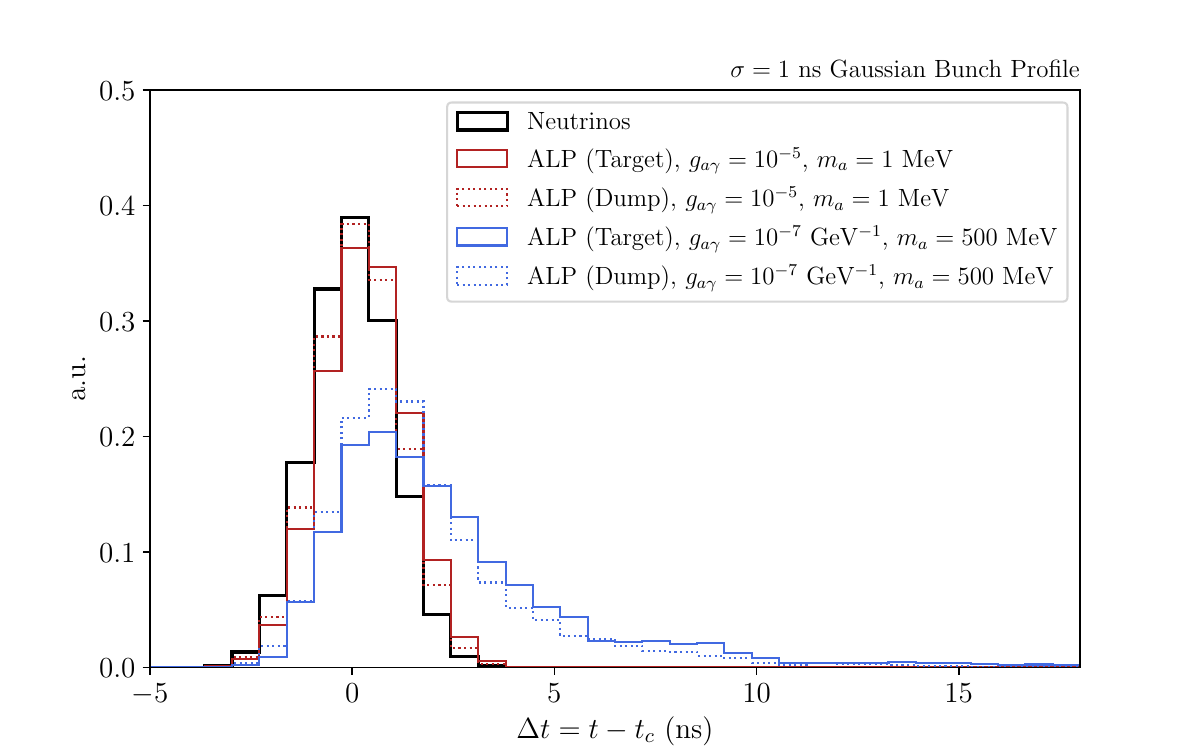}
    \caption{Timing spectra are shown for several benchmark masses and ALP-photon couplings. }
    \label{fig:timing}
\end{figure}

The timing distribution of long-lived particles as they arrive at the near detector offers another potential handle to distinguish signal from background. For ALPs at DUNE, the masses of interest can be as large as 1 GeV, with a variety of decay lifetimes that can make the ALP more or less boosted in the laboratory frame. This means that the time of flight from the source of ALP production to the detector can vary depending on the ALP mass and its lifetime, sometimes manifesting in an arrival time distribution that is significantly different from the neutrino background.

In addition, contributions to the ALP production at the LBNL beam dump will leave an additional imprint on the timing signature; the total time of flight difference will now be comprised of the proton travel time from the target to the dump in addition to the ALP time of flight from the dump to the detector, relative to the neutrino total time of flight ($\approx 1.9 \cdot 10^{-6}$ $\mu$s). These effects and others have been investigated in ref.~\cite{Dutta:2025npn} in the context of heavy neutral leptons at the SBND, MicroBooNE, and ICARUS experiments; similar qualities in the timing spectra should be present in the ALP case we consider in this work.

In Fig.~\ref{fig:timing}, we show some example timing spectra of the ALP flux arriving at the near detector. These are computed using a Gaussian-distributed beam spill approximation as a toy model for the distribution of production times, $t_g \sim G(0, \sigma)$ where we take $\sigma = 1$ nanosecond. The true distribution of the neutrino timing signal may be further smeared by meson lifetimes and neutrino production distributed between the target, focusing horns, and decay pipe. The 1 ns timing structure is motivated by refs.~\cite{DUNE:2015lol,Angelico:2019gyi} with ND-LAr timing capabilities motivated in refs.~\cite{Paudel:2023ove,MicroBooNE:2023ldj}. As we do not yet know the exact description of the timing spectra, we did not include it in the analysis to establish sensitivities to the ALP signal; however, Fig.~\ref{fig:timing} does motivate the possibility of including the arrival time distribution to distinguish ALP or other long-lived particle signals from the neutrino signal. We leave a more sophisticated treatment of timing distributions as a background rejection handle to future work.

\bibliography{main}

\end{document}